\documentclass{elsart}

\usepackage{natbib}

\usepackage{amssymb}

\begin{document}

\begin{frontmatter}

\title{Pascual Jordan's resolution of the conundrum \\ of the wave-particle
duality of light.\thanksref{mpiwg}}
\thanks[mpiwg]{This paper was written as part of a joint project in the history of quantum physics of the {\it Max Planck Institut f\"{u}r Wissenschaftsgeschichte} and the {\it Fritz-Haber-Institut} in Berlin.}

\author[duncan]{Anthony Duncan},
\author[janssen]{Michel Janssen\corauthref{cor}}
\corauth[cor]{Corresponding author. Address: Tate Laboratory of Physics, 116 Church St.\ SE, Minneapolis, MN 55455, USA, Email: janss011@umn.edu}
\address[duncan]{Department of Physics and Astronomy, University of Pittsburgh}
\address[janssen]{Program in the History of Science, Technology, and Medicine, \\ University of Minnesota}

\begin{abstract}
In 1909, Einstein derived a formula for the mean square energy fluctuation in a subvolume of a box filled with black-body radiation. This formula is the sum of a wave term and a particle term. Einstein concluded that a satisfactory theory of light would have to combine aspects of a wave theory and a particle theory. In a key contribution to the 1925 {\it Dreim\"annerarbeit} with Born and Heisenberg, Jordan used Heisenberg's new {\it Umdeutung} procedure to quantize the modes of a string and argued that in a small segment of the string, a simple model for a subvolume of a box with black-body radiation, the mean square energy fluctuation is described by a formula of the form given by Einstein. The two terms thus no longer require separate mechanisms, one involving particles and one involving waves. Both terms arise from a single consistent dynamical framework. Jordan's derivation was later criticized by Heisenberg and others and the lingering impression in the subsequent literature is that infinities of one kind or another largely invalidate Jordan's conclusions. In this paper, we carefully reconstruct Jordan's derivation and reexamine some of the objections raised against it. Jordan could certainly have presented his argument more clearly. His notation, for instance, fails to bring out that various sums over modes of the string need to be restricted to a finite frequency range for the final result to be finite. His derivation is also incomplete. In modern terms, Jordan only calculated the quantum uncertainty in the energy of a segment of the string in an eigenstate of the Hamiltonian for the string as a whole, while what is needed is the spread of this quantity in a thermal ensemble of such states. These problems, however, are easily fixed. Our overall conclusion then is that Jordan's argument is basically sound and that he deserves credit for resolving a major conundrum in the development of quantum physics.
\end{abstract}


\end{frontmatter}

\section{The recovery of Einstein's fluctuation formula in the {\it Drei\-m\"anner\-arbeit}}

In the final section of the famous {\it Dreim\"annerarbeit} of Max Born, Werner Heisenberg, and Pascual Jordan (1926)---cited hereafter as `3M'---the {\it Umdeutung} [= reinterpretation] procedure of \citep{Heisenberg 1925} is applied to a simple system with infinitely many degrees of freedom, a continuous string fixed at both ends. In a lecture in G\"ottingen in the summer of 1925 (3M, p.\ 380, note 2)---attended, it seems, by all three authors of the  {\it Dreim\"annerarbeit}---Paul \citet{Ehrenfest 1925a} had used this system as a one-dimensional model for a box filled with black-body radiation and had calculated the mean square energy fluctuation in a small segment of it. The string can be replaced by a denumerably infinite set of uncoupled harmonic oscillators, one for every possible mode of the string. The harmonic oscillator is the simplest application of Heisenberg's new quantum-theoretical scheme. The basic idea behind this scheme was to retain the classical equations of motion but to {\it reinterpret} them---hence the term {\it Umdeutung}---as expressing relations between arrays of numbers, soon to be recognized as matrices \citep{Born and Jordan 1925}, assigned not to individual states but to transitions between states and subject to a non-commutative multiplication law.\footnote{See \citep{Duncan and Janssen} both for an account of what led Heisenberg to this idea and for further references to the extensive historical literature on this subject. Both our earlier paper and this one are built around the detailed reconstruction and elementary exposition of one key result---Van Vleck's derivation of the Kramers dispersion formula in the former (secs.\ 5.1--5.2 and 6.2) and the derivation of the mean square energy fluctuation in the small segment of a string in the {\it Dreim\"annerarbeit} in the latter (sec.\ 3 below).} When this {\it Umdeutung} procedure is applied to the harmonic oscillators representing the modes of a string and the mean square energy fluctuation in a small segment of the string and a narrow frequency interval is calculated, one arrives at a surprising result. In addition to the classical wave term, proportional to the square of the mean energy, one finds a term proportional to the mean energy itself. This term is just what one would expect  for the mean square energy fluctuation in a system of particles.

For this simple model, one thus recovers both terms of the well-known formula for the mean square energy fluctuation in a subvolume of a box with black-body radiation that Albert Einstein showed in 1909 is required by the Planck formula for black-body radiation and some general results in statistical mechanics. After deriving a similar formula for the mean square fluctuation of momentum, Einstein remarked that in both cases ``the effects of the two causes of fluctuation mentioned [waves and particles]  act like fluctuations (errors) arising from mutually independent causes (additivity of the terms of which the square of the fluctuation is composed)" \citep[p.\ 190]{Einstein 1909a}. He  reiterated this point in his lecture later that year at a meeting of the {\it Gesellschaft Deutscher Naturforscher und \"Arzte} in Salzburg \citep[p.\ 498]{Einstein 1909b}.\footnote{See also \citep[p.\ 346]{Einstein 1914}.} To paraphrase a modern commentator, black-body radiation seems to behave as a combination of waves and particles acting independently of one another \citep[p. 178]{Bach 1989}. These fluctuation formulae famously led Einstein to predict at the beginning of his Salzburg lecture ``that the next phase of the development of theoretical physics will bring us a theory of light that can be interpreted as a kind of fusion of the wave and emission theories" \citep[pp.\ 482--483]{Einstein 1909b}.\footnote{For recent discussions of Einstein's early use of fluctuation considerations to advance his light-quantum hypothesis, see, e.g., the contributions of  J\"urgen Renn and Robert Rynasiewicz (2006: Ryno-Renn I), John \citet{Norton 2006}, and Jos \citet{Uffink 2006} to a special issue of {\it Studies in History and Philosophy of Modern Physics} on the centenary of Einstein's {\it annus mirabilis}.}  

Reluctant to abandon the classical theory of electromagnetic radiation and to embrace Einstein's light-quantum hypothesis, several physicists in the following decade tried to find either holes in Einstein's derivation of the (energy) fluctuation formula so that they could avoid the formula or an alternative derivation of it with which they could avoid light quanta.\footnote{\label{kojevnikov}
For historical discussion and references to both primary and secondary literature on this issue, see 
\citep{Bach 1989} and \citep{Kojevnikov 1990}. For some brief comments, see \citep[p.\ 642, note 2]{Jordan 1927b}, \citep[p.\ 398, note 1]{Born and Jordan 1930}, and \citep[p.\ 220]{Jordan 1936}.} In the wake of the Compton effect \citep{Compton 1923} and Satyendra Nath Bose's new derivation of Planck's black-body radiation law \citep{Bose 1924}, however, both Einstein's fluctuation formula and his light quanta began to look more and more unavoidable. As a result, the problem of reconciling the wave and the particle aspects of light took on greater urgency. In the paper that provided the simple model used in the {\it Dreim\"annerarbeit}, for instance, \citet{Ehrenfest 1925a} emphasized the paradoxical situation that quantizing the modes of a classical wave according to a method proposed by Peter \citet{Debye 1910} gives the correct Planck formula for the energy distribution  in black-body radiation but the wrong formula for the mean square fluctuation of the energy \citep[p.\ 379]{Stachel 1986}. This problem is also highlighted in the {\it Dreim\"annerarbeit} (3M, p.\ 376). As Einstein characterized the situation in an article on the Compton effect in the  {\it Berliner Tageblatt} of April 20, 1924:  ``There are \ldots now two theories of light, both indispensable and---as one must admit today despite twenty years of tremendous effort on the part of theoretical physicists---without any logical connection."\footnote{This often-quoted passage can be found, for instance, in (Pais, 1980, 211; 1982, p.\ 414; 1986, p.\ 248), \citep[p.\ 182]{Klein 1980}, and \citep[p.\ 182]{Bach 1989}.} One possibility that was seriously considered at the time, especially after the decisive refutation  in April 1925 of the theory of \citet{BKS}, which can be seen as a last-ditch effort to avoid light quanta, was that light consisted of particles guided by waves \citep[sec.\ 4.2]{Duncan and Janssen}. In this picture, the waves and the particles presumably give separate contributions to the mean square energy fluctuation, just as Einstein had envisioned in 1909. A derivation of Einstein's fluctuation formula based on this picture (and Bose statistics) was given by Walther \citet{Bothe 1927}.\footnote{Two earlier papers by \citet{Bothe 1923, Bothe 1924} and a related paper by Mieczyslaw \citet{Wolfke 1921} are cited in the {\it Dreim\"annerarbeit} (3M, p.\ 379, notes 2 and 3).} The {\it Dreim\"annerarbeit} was published February 4, 1926, but Bothe's paper, submitted in December 1926, makes no reference to it.
The derivation in the {\it Dreim\"annerarbeit} shows, contrary to what Bothe suggested and contrary to what Einstein expected, that the two terms in the fluctuation formula do not require separate mechanisms after all but can be accounted for within a single consistent dynamical framework. 

The authors presented their unified mechanism in terms of (quantized) waves, but it can also be described in terms of (quantum) particles. \citet{Heisenberg 1930} made this quite explicit a few years later:  ``The quantum theory, which one can interpret as a particle theory or a wave theory as one sees fit, leads to the complete fluctuation formula" (p.\ 101; see also, e.g., Jordan, 1936, p.\ 220).\footnote{\label{fuerth}Independently of full-fledged quantum mechanics and using only Bose's quantum statistics, Reinhold \citet[p.\ 312]{Fuerth 1928}, a theoretical physicist in Prague, argued that the fluctuation formula was compatible with waves, particles, or a combination of both. After fleeing Czechoslovakia in 1938, F\"urth ended up in Edinburgh, where he collaborated with Born. In his memoirs, \citet{Born 1978} calls him ``a scholar of the highest rank," who ``had worked mainly in the field of statistical mechanics, in particular about thermal fluctuations" (p.\ 289).} The result thus illustrates the kind of wave-particle duality associated with Bohr's complementarity principle. This kind of wave-particle duality is different from that originally envisioned by Einstein; it does not posit the coexistence of two different mechanisms but the existence of one that can be described in different ways. While illustrating one aspect of complementarity, the fluctuation formula undermines another.  Complementarity also involves the notion that  a quantum system, depending on the experimental context,  presents itself to us either under the guise of waves or under the guise of particles. However, if we were to measure the mean square energy fluctuation in a small subvolume of a box with black-body radiation, which in principle we could even though Einstein conceived of it only as a thought experiment, and Einstein's formula is correct, which is no longer in any serious doubt,\footnote{{\it Pace} \citep{Gonzalez and Wergeland 1973}.} we would see the effects of waves and particles 
simultaneously.\footnote{We owe this last observation to Jos Uffink (private communication).}   

One might have expected that the recovery of Einstein's fluctuation formula in the {\it Dreim\"annerarbeit} would immediately have been heralded as the triumphant resolution of a major conundrum in the development of quantum physics; and that it would since have become a staple of historical accounts of the wave-particle duality of light. Both expectations prove to be wrong. We already encountered two early skeptics, Bothe and F\"urth (see note \ref{fuerth}), and we shall shortly encounter more prominent ones (see sec.\ 2). To give some counterexamples from the historical literature, Martin J.\ Klein's (1964) classic ``Einstein and the wave-particle duality," far from making the {\it Dreim\"annerarbeit} the natural denouement of the narrative, does not even cite the paper;\footnote{The same is true for Klein's (1979, 1980, 1982) contributions to three volumes published in connection with the centenary of Einstein's birth, even though the first briefly touches on Einstein's reaction to matrix mechanics \citep[p.\ 149]{Klein 1979} and the third is specifically on Einstein and fluctuations \citep{Klein 1982}. In a much earlier paper on Ehrenfest, \citet[p.\ 50]{Klein 1959} mentioned the importance of \citep{Ehrenfest 1925a} for this part of the {\it Dreim\"annerarbeit}, adding that ``it takes some careful quantum mechanical arguments to give a satisfactory discussion of the ``mechanism" of the fluctuations" and that the ``situation was not clarified until many years later." Klein cites \citep{Heisenberg 1931} as providing that clarification. In our estimation, with one important {\it caveat} that has nothing to do with Heisenberg's criticism (see sec.\ 3), the argument in the {\it Dreim\"annerarbeit} is fine, although it certainly could have been stated more clearly. Klein, incidentally, was the communicating editor for  a paper by Alexander \citet{Bach 1989}, who is very dismissive of the treatment of fluctuations in the {\it Dreim\"annerarbeit} (see note \ref{abelian}).} nor is it mentioned in a recent paper on the history of the light-quantum hypothesis by Stephen G.\ \citet{Brush 2007}.\footnote{We should add that Brush is mainly concerned with experimental developments, makes only a passing reference to Einstein's 1909 work (ibid., p.\ 233), and does not cite \citep{Einstein 1909a, Einstein 1909b},  which first posed the puzzle addressed in the {\it Dreim\"annerarbeit}.} To the best of our knowledge, the only Einstein biography that even touches on the derivation of the fluctuation formula in the {\it Dreim\"annerarbeit} is the one by Abraham \citet[p.\ 405]{Pais 1982}.\footnote{It is not mentioned in \citep{Pais 1979} or \citep{Pais 1980}, however, which served as the basis for this part of Pais's biography of Einstein. Pais added these comments as ``a postscript" to a section on the ``fusion of particles and waves" and refers to \citep{Gonzalez and Wergeland 1973} for further discussion \citep[pp.\ 402--405]{Pais 1982}.} 
The canonical twin stories of the light-quantum hypothesis and the wave-particle duality of light end with the Compton effect and Bohr complementarity, respectively.\footnote{\citet[p.\ 237]{Brush 2007}, for instance, is referring to complementarity when he lists the realization that ``an entity can behave like a wave motion in some experiments and like a stream of particles in another" as one of reasons for the acceptance of the light-quantum hypothesis.} The canonical history fails to mention that the specific challenge posed by Einstein's fluctuation formula, which suggested wave-particle duality in the first place, was in fact taken up and, we want to argue, convincingly met in the {\it Dreim\"annerarbeit}.\footnote{Another episode in the history of Einstein and wave-particle duality that seldom gets attention is the one involving the fraudulent canal ray experiments of Emil Rupp (see Van Dongen, 2007).}  It also tends to ignore the difference that we drew attention to above between Einstein's original conception of wave-particle duality and wave-particle duality as it is usually understood in quantum mechanics.\footnote{\label{pais}E.g., after quoting Einstein's 1909 prediction about the ``fusion of the wave and emission theories," \citet{Pais 1982} comments: ``This fusion now goes by the name of complementarity" (p.\ 404).} 

Having chastised historians of physics in such broad-brush fashion, we hasten to add that ours is hardly the first contribution to the historical literature to draw attention to this part of the {\it Dreim\"annerarbeit} nor the first to give it its due. For instance, even though \citet{Klein 1980} did not mention the fluctuation calculation in the {\it Dreim\"annerarbeit} in his lecture at the Princeton Einstein centenary symposium, John Stachel, director of the Einstein Papers Project at the time, did bring it up in question time \citep[p.\ 196]{Woolf 1980}. Stachel also alerted the audience to  correspondence between Einstein and Jordan pertaining to this calculation, much of which, unfortunately, is no longer extant. Jagdish Mehra and Helmut Rechenberg (1982--2001, Vol.\ 3, pp.\ 149--156) devote a section of their comprehensive history of quantum mechanics to this part of the {\it Dreim\"annerarbeit}, although they offer little assistance to a reader having difficulties following the argument in the original paper. The derivation of Einstein's fluctuation formula in the  {\it Dreim\"annerarbeit} also plays a central role in a paper on Einstein's fluctuation formula and wave-particle duality by Alexei \citet{Kojevnikov 1990} (cf.\ note \ref{kojevnikov}); in a recent paper on Jordan's contributions to quantum mechanics by J\"urgen \citet{Ehlers 2007}; and in a paper on the origin of quantized matter waves by Olivier \citet{Darrigol 1986}. This last author clearly shares our enthusiasm for these fluctuation calculations, calling them ``spectacular" at one point and stressing that they formed the solution to ``the most famous puzzle of radiation theory" \citep[p.\ 221--222]{Darrigol 1986}. 

The second half of our paper (sec.\ 3) will be given over to a detailed and self-contained reconstruction of the calculation in the {\it Dreim\"annerarbeit}  of the mean square energy fluctuation in a small segment of a string, Ehrenfest's simple one-dimensional model for a subvolume of a box filled with black-body radiation. This reconstruction will enable us to defend the argument against various criticisms that have been raised against it and will enable our readers to assess such criticisms and our rejoinders for themselves. 
In the first half of our paper (sec.\ 2), drawing heavily on the work of Darrigol and Kojevnikov cited above, we want to explore the question why the recovery of Einstein's fluctuation formula in the {\it Dreim\"annerarbeit} is not nearly as celebrated as one would expect it to be. In sec.\ 4 we briefly summarize our conclusions. 

\section{Why is the solution to Einstein's riddle of the wave-particle duality of light not nearly as famous as the riddle itself?}  

\subsection{One of Jordan's most important contributions to physics}

The first thing to note in the search for answers to the question posed in the title of this section is  
that the derivation of the fluctuation formula in the {\it Dreim\"annerarbeit}, though presented as part of a collaborative effort, was actually the work of only one of the authors, namely Jordan, ``the unsung hero among the creators of quantum mechanics" \citep[p.\ 5]{Schweber 1994}. Today, Jordan is mostly remembered as perhaps the only first-tier theoretical physicist  who sympathized strongly and openly with the Nazi ideology.\footnote{\label{nazi}For a detailed recent discussion of this aspect of Jordan's life and career, see \citep{Hoffmann and Walker 2007}.} It is hard to say whether this entanglement has been a factor in the neglect of the derivation of the fluctuation formula in the {\it Dreim\"annerarbeit}.
Our impression, for what it is worth, is that it only played a minor role.
It should be recalled in this context that it was not until 1930 that Jordan began to voice his Nazi sympathies in print and then only under the pseudonym of Ernst Domaier \citep[p.\ 71]{Beyler 2007}. A much more important factor, it seems, was that Jordan's result immediately met with resistance, even from his co-authors and from the person responsible for the riddle he had solved. 
Right from the start a cloud of  suspicion surrounded the result and that cloud never seems to have lifted.\footnote{For criticism see, e.g., \citep{Smekal 1926}, \citep{Heisenberg 1931}, \citep{Born and Fuchs 1939a}, \citep{Gonzalez and Wergeland 1973}, and \citep{Bach 1989}.}  Our paper can be seen as an attempt to disperse it.

Except for a short period of wavering in 1926, Jordan steadfastly stood by his result and considered it one of his most important contributions to quantum mechanics. He said so on a number of occasions. One such occasion was  a conference honoring Paul A.\ M.\ Dirac on his 70th birthday. At the conference itself, Jordan talked about ``the expanding earth" \citep[p.\ 822]{Mehra 1973}, a topic that apparently occupied him for 20 years.\footnote{\citep[p.\ 124]{Kundt 2007}. Jordan argued that such an expansion would result if the gravitational coupling constant $G(t)$ were decreasing over time and that this might help explain continental drift.}
For the proceedings volume, however, he submitted some reminiscences about the early years of quantum mechanics. There he wrote: 
\begin{quotation}
Another piece in the `Dreim\"anner Arbeit' gave a result, which I myself have been quite proud of: It was possible to show that the laws of fluctuations in a field of waves, from which Einstein derived the justification of the concept of corpuscular light quanta, can be understood also as consequences of an application of quantum mechanics to the wave field \citep[p.\ 296]{Jordan 1973}.
\end{quotation}
In the early 1960s, Jordan had likewise told  Bartel L. Van der Waerden that he ``was very proud of this result at the time," adding that he ``did not meet with much approval."\footnote{\label{letter120161}Jordan to Van der Waerden, December 1, 1961. Transcriptions of correspondence between Jordan and Van der Waerden in 1961--1962 can be found in the folder on Jordan in the {\it Archive for History of Quantum Physics}, cited hereafter as AHQP \citep{Kuhn et al. 1967}. We consulted the copy of this archive at Walter Library, University of Minnesota. Van der Waerden relied heavily on his correspondence with Jordan in editing his well-known anthology \citep{Van der Waerden}.} In a follow-up letter, Jordan wrote:
\begin{quotation}
What [Born and Jordan 1925]  says about radiation is not very profound. But what the {\it Dreim\"annerarbeit} says about energy fluctuations in a field of quantized waves is, in my opinion, {\it almost the most important contribution I ever made to quantum mechanics}.\footnote{\label{letter041062}``\ldots nahezu das Wichtigste, was ich \"uberhaupt zur Quantenmechanik beigetragen habe." Jordan to Van der Waerden, April 10, 1962 (AHQP), our emphasis. In view of the first sentence, it is not surprising that  chapter 4 of \citep{Born and Jordan 1925}, ``comments on electrodynamics," was left out of \citep{Van der Waerden}.}
\end{quotation}
J\"urgen \citet{Ehlers 2007}, who studied with Jordan, relates: ``In the years that I knew him, Jordan rarely talked about his early work. On a few occasions, however, he did tell me that he was especially proud of having derived Einstein's fluctuation formula \ldots by quantizing a field" (p.\ 28).

In late 1925, when the {\it Dreim\"annerarbeit} was taking shape, Jordan was probably the only physicist  who had done serious work both on the light-quantum hypothesis and on the new matrix mechanics. In his dissertation, supervised by Born and published as \citep{Jordan 1924}, he had criticized the argument for ascribing momentum to light quanta in \citep{Einstein 1917}. He had ``renounced this heresy" ({\it jede Ketzerei aufgegeben})\footnote{P.\ 13 of the transcript of session 2 of 
Thomas S.\ Kuhn's 
 interview with Jordan in June 1963 for the AHQP.
For further discussion of \citep{Jordan 1924, Jordan 1925}, see session 1, pp.\ 10--11, 15, and session 2, pp.\ 16--17 of the interview. For discussion of the section on fluctuations in the {\it Dreim\"annerarbeit}, including Jordan's views of the work by \citet{Bothe 1923, Bothe 1924}, see session 3, pp.\ 8--9.}  after \citet{Einstein 1925} published a brief rejoinder. A paper published the following year, \citep{Jordan 1925}, shows that its author was up on the latest statistical arguments concerning light quanta. Later in 1925 Jordan gave Born a manuscript for publication in {\it Zeitschrift f\"ur Physik} in which he essentially proposed what is now known as Fermi-Dirac statistics. Unfortunately, the manuscript ended up at the bottom of a suitcase that Born took with him to the United States and did not resurface until Born returned from his trip, at which point Jordan had been scooped \citep[p.\ 49]{Schroer 2007}. In addition to his work in quantum statistics, Jordan was one of the founding fathers of matrix mechanics. The {\it Dreim\"annerarbeit} is the sequel to \citep{Born and Jordan 1925}, which greatly clarified the theory proposed in Heisenberg's {\it Umdeutung} paper. Unfortunately for Jordan, few if any physicists at the time were primed for his sophisticated combination of these two contentious lines of research---the statistics of light quanta and matrix mechanics---in his derivation of Einstein's fluctuation formula.

\subsection{The reactions of Heisenberg and Born to Jordan's result}

As we mentioned, even Jordan's co-authors experienced great difficulty understanding his reasoning and entertained serious doubts about its validity. In the letter cited in note \ref{letter120161}, Jordan gave Van der Waerden a breakdown of who did what in the  {\it Dreim\"annerarbeit} and commented that ``my reduction of light quanta to quantum mechanics was considered misguided [{\it abwegig}] by Born and Heisenberg for a considerable period of time." In the letter to Van der Waerden cited in note \ref{letter041062}, Jordan, after  reiterating that these fluctuation considerations were ``completely mine" ({\it ganz von mir}),  elaborated on the resistance he encountered from his co-authors
\begin{quotation}
Later, Heisenberg in fact explicitly questioned whether this application I had made of quantum mechanics to a system of {\it infinitely many degrees of freedom} was correct. It is true that Born did not second Heisenberg's opinion at the time that it was wrong, but he did not explicitly reject Heisenberg's negative verdict either.\footnote{\label{letter041062a}Jordan to Van der Waerden, April 10, 1962 (AHQP).} 
\end{quotation}
Jordan's recollections fit with statements made by his co-authors at various points in time. 

A few weeks before the {\it Dreim\"annerarbeit} was submitted to {\it Zeitschrift f\"ur Physik}, where it was received November 16, 1925, Heisenberg wrote to Wolfgang Pauli: ``Jordan claims that the interference calculations come out right, both the classical [wave] and the Einsteinian [particle] terms \ldots I am a little unhappy about it, because I do not understand enough statistics to judge whether it makes sense; but I cannot criticize it either, because the problem itself and the calculations look  meaningful to me."\footnote{Heisenberg to Pauli, October 23, 1925  \citep[p.\ 252]{Pauli 1979}, quoted (in slightly different translations) and discussed in \citep[p.\ 220]{Darrigol 1986} and in \citep[Vol.\ 3, p.\ 149]{Mehra Rechenberg}.} This is hardly a ringing endorsement. By 1929 Heisenberg had warmed to Jordan's fluctuation calculations and he included them in the published version of lectures on quantum mechanics that year at the University of Chicago \citep[Ch.\ V, sec.\ 7]{Heisenberg 1930}.\footnote{We already quoted Heisenberg's conclusion in the introduction.} On the face of it, \citet{Heisenberg 1931} lost faith again the following year, when he showed that  the mean square energy fluctuation diverges {\it if we include all possible frequencies}, even in a model that avoids the zero-point energy of the harmonic oscillators of the model used in the {\it Dreim\"annerarbeit}. As we shall see in sec.\ 3, however, Jordan clearly intended to calculate the mean square fluctuation of the energy {\it in a narrow frequency range}, even though his notation in various places suggests otherwise. In that case, the mean square fluctuation is perfectly finite, regardless of whether we consider a model with or without zero-point energy. It is true that the mean square energy fluctuation as calculated by Jordan diverges when integrated over all frequencies. As Heisenberg showed in his 1931 paper, however, this is essentially an artifact of the idealization that the subvolume for which the energy fluctuations are computed has sharp edges. If we smooth out the edges, the mean square energy fluctuation remains finite even when integrated over all frequencies.\footnote{See the discussion following Eq.\ (\ref{q24}) in sec.\ 3.2 for further details on Heisenberg's objection and its resolution.}

Jordan included his fluctuation argument in the book on matrix mechanics co-authored with Born and conceived of as a sequel to \citep{Born 1925} \citep[sec.\ 73, pp.\ 392--400]{Born and Jordan 1930}. By the early 1930s, anyone who cared to know must have known that Jordan was responsible for this part of the {\it Dreim\"annerarbeit}. This can be inferred, for instance, from Pauli's scathing review of Born and Jordan's book, in which he famously praises only the typesetting and the quality of the paper. The reviewer wearily informs his readers that the authors once again trot out the ``trains of thought about fluctuation phenomena, which one of the authors (P.\ Jordan) has already taken occasion to present several times before" \citep{Pauli 1930}.\footnote{For what it is worth, we note that Jordan's result is not mentioned in Pauli's (1949) overview of Einstein's contributions to quantum theory in the Schilpp volume, even though the fluctuation formula is mentioned prominently.}
These considerations were indeed included, for instance, in two semi-popular articles by Jordan on recent developments in quantum physics, first in \citep[p.\ 642]{Jordan 1927b} and then, more extensively, in a fifty-page history of the light-quantum hypothesis \citep[sec.\ 13, pp.\ 192--196]{Jordan 1928}. They 
also form the starting point of Jordan's review of the current state of quantum electrodynamics at a conference in Charkow the following year \citep[pp.\ 700-702]{Jordan 1929}. We shall have occasion to quote from these texts below.

That these fluctuation considerations make yet another appearance in \citep{Born and Jordan 1930} would, if nothing else, seem to indicate Born's (belated) approval of Jordan's argument. In the late 1930s, however, in a paper written in exile in Edinburgh with  Klaus Fuchs, his assistant at the time,\footnote{For Born's reaction to Fuchs's later arrest as a Soviet spy, see \citep[p.\ 288]{Born 1978}.}
Born sharply criticizes Jordan's argument as well as Heisenberg's 1931 amendment to it. He goes as far as dismissing a central step in the argument in the {\it Dreim\"annerarbeit} as ``quite incomprehensible reasoning," offering as his  only excuse for signing off on this part of the paper ``the enormous stress under which we worked in those exciting first days of quantum mechanics" \citep[p.\ 263]{Born and Fuchs 1939a}.\footnote{\label{born complaint}In our reconstruction  of Jordan's argument in sec.\ 3, we shall identify the step that Born and Fuchs found so objectionable (see note \ref{incomprehensible}).} 

Born conveniently forgets to mention that he had signed his name to the same ``incomprehensible reasoning" in his 1930 book with Jordan. If someone had reminded him, however, he probably would have invoked stress again. In the fall of 1930, still smarting from Pauli's sarcastic review, he wrote to Arnold Sommerfeld, Pauli's teacher:
\begin{quotation}
[B]ecause I think very highly of Pauli's accomplishments, I am sorry that our personal relationship is not particularly good. You will probably have realized that on the basis of his criticism of  [Born and Jordan, 1930] in the {\it Naturwissenschaften}. I know full well that the book has major weaknesses, which are due in part to the fact that it was started too early and in part to the fact that I fell ill during the work, a collapse from which unfortunately I still have not fully recovered. But from Pauli's side the nastiness of the attack has other grounds, which are not very pretty.\footnote{Born to Sommerfeld, October 1, 1930, quoted in Von Meyenn, 2007, pp.\ 45-47.}
\end{quotation}
Born then explains to Sommerfeld how he had originally asked Pauli to collaborate with him on the development of matrix mechanics and how he had only turned to Jordan after Pauli had turned him down (see Born, 1978, pp.\ 218--219, for the canonical version of this story). ``Ever since Pauli took himself out in this manner," Born's letter to Sommerfeld continues, ``he has had a towering rage against G\"ottingen and has wasted no opportunity to vent it through mean-spirited comments" (Born to Sommerfeld, October 1, 1930). Born eventually came to agree with the substance of Pauli's criticism.
In his memoirs, in a chapter written in the early 1960s  \citep[p.\ 225]{Born 1978}, Born is very dismissive of his book with Jordan and conceded that the authors' self-imposed restriction to matrix methods was a ``blunder" for which they had rightfully been excoriated by Pauli. As Pauli had put it in his review:
\begin{quotation}
Many results of quantum theory \ldots cannot be derived at all [with the techniques presented in Born and Jordan's book], others only very inconveniently and through indirect methods (to the latter category, for instance, belongs the derivation of the Balmer terms, which is carried out matrix-mechanically following an earlier paper on the topic by Pauli [1926]. One therefore cannot reproach the reviewer that he only finds the grapes sour because they are hanging too high for him) \citep{Pauli 1930}.
\end{quotation}
In his memoirs, Born's ire is directed not at Pauli but at Jordan, whom he blames for the G\"ottingen parochialism---or ``local patriotism," as he calls it---that led them to use matrix methods only \citep[p.\ 230]{Born 1978}. It is possible that Born had already arrived at this assessment when he attacked Jordan's fluctuation considerations in his paper with Fuchs. 

Whatever the case may be, a few months after the publication of their paper, \citet{Born and Fuchs 1939b} had to issue a ``correction."   Markus E.\ Fierz, Pauli's assistant in Zurich at the time, had alerted them to a serious error in their calculations. The resulting two-page ``correction" amounts to a wholesale retraction of the original paper. The authors explicitly withdraw their criticism of \citep{Heisenberg 1931} but do not extend the same courtesy to Jordan. This same pattern returns in Born's memoirs, in another chapter dating from the early 1960s, where Born writes that he and Fuchs ``worked on the fluctuations in the black-body radiation but discovered later that Heisenberg had done the same, and better" \citep[p.\ 285]{Born 1978}. We find it hard to suppress the thought that, starting sometime in the 1930s, Born's perception of Jordan and Jordan's work became colored---and who can blame him?---by his former student's manifest Nazi sympathies. 

\subsection{Jordan's result and the birth of quantum field theory}

As Jordan emphasized, both in the late 1920s and reflecting on this period later, behind the initial resistance of Born and Heisenberg to his fluctuation calculation was a more general resistance to the notion of quantizing the electromagnetic field. As he told Kuhn: ``The idea that from the wave field, i.e., from the electromagnetic field, one had to take another step to quantum mechanics struck many physicists back then as too revolutionary, or too artificial, or too complicated, and they would rather not believe it."\footnote{AHQP interview with Jordan, session 3, p.\ 8.} The passage from a letter from Jordan to Van der Waerden (see note \ref{letter041062a}) that we quoted above already hints at this and it is made more explicit as the letter continues:
\begin{quotation}
I remember that, to the extent that they took notice of these issues at all, other theorists in G\"ottingen [i.e., besides Born and Heisenberg], [Yakov] Frenkel for instance,  considered my opinion, expressed often in conversation, that the electromagnetic field and the Schr\"odinger field had to be quantized \ldots  as a somewhat fanciful exaggeration or as lunacy.\footnote{``\ldots eine etwas phantasti\-sche Uebertreibung oder Verr\"ucktheit."} This changed only when Dirac [1927] also quantized both the electromagnetic field and the field of matter-waves. I still remember how Born, who had been the first to receive an offprint of the relevant paper of Dirac, showed it to me and initially looked at it shaking his head. When I then pointed out to him that I had been preaching the same idea all along ever since our {\it Dreim\"annerarbeit}, he first acted surprised but then agreed.\footnote{Jordan's text can be read as saying that Born agreed that Jordan had indeed been championing the same idea, but what he meant, presumably, is that Dirac's paper convinced Born of the merit of the idea.} Heisenberg then also set aside his temporary skepticism, though it was not until considerably later that he himself started to work toward a quantum theory of fields (or ``quantum electrodynamics") in the paper he then published with Pauli\footnote{\citep{Heisenberg and Pauli 1929, Heisenberg and Pauli 1930}}, which followed up on my three joint papers with Pauli, [Oskar] Klein, and [Eugene] Wigner\footnote{\citep{Jordan and Pauli 1928, Jordan and Klein 1927, Jordan and Wigner 1928}. }$^,$\footnote{Jordan to Van der Waerden, April 10, 1962 (AHQP).} 
\end{quotation}
These and other publications of the late 1920s make Jordan one of the pioneers of quantum field theory. In his AHQP interview, Jordan told Kuhn the same story he told Van der Waerden.\footnote{Unfortunately, Jordan either misspoke at one point or there is an error in the transcription: ``Born also did not want to know anything about [quantizing the Schr\"odinger wave function] until one day the paper by Dirac arrived in which this was done. At the time he was surprised and wondered about it, and I said: ``Yes, that is exactly what you [{\it Sie}; this should clearly be `I'] said: it should be possible to do that." ``Yes, of course!" [({\it Ach richtig}!) said Born]" (AHQP interview with Jordan, session 3, p.\ 9).} Kuhn asked him in this context who had coined the phrase ``second quantization." Jordan told him he had (ibid.).  A version of the story he told Van der Waerden and Kuhn in the early 1960s can already be found in a letter to Born of the late 1940s:
\begin{quotation}
It has always saddened me somehow that the attack on the light-quantum problem already contained in our {\it Dreim\"annerarbeit} was rejected by everyone for so long (I vividly remember how Frenkel, despite his very friendly disposition toward me, regarded the quantization of the electromagnetic field as a mild form of insanity\footnote{``\ldots eine Art leichtes Irresein" [sic].}) until Dirac took up the idea {\it from which point onward he was the only one cited in this connection}.\footnote{Jordan to Born, July 3, 1948 (AHQP), our emphasis.}
\end{quotation}
Given the resentment one senses in the italicized clause, there is some irony in how Jordan segues into the more detailed (and hence, one suspects, somewhat embellished) version of this same story in his reminiscences for the volume in honor of Dirac's 70th birthday:
\begin{quotation}
{\it I have been extremely thankful to Dirac in another connection}. My idea that the solution of the vexing problem of Einstein's light quanta might be given by applying quantum mechanics to the Maxwell field itself, aroused the doubt, scepticism, and criticism of several good friends. But one day when I visited Born, he was reading a new publication of Dirac, and he said: `Look here, what Mr.\ Dirac does now. He assumes the eigenfunctions of a particle to be non-commutative observables again.' I said: `Naturally.' And Born said: `How can you say ``naturally"?' I said: `Yes, that is, as I have asserted repeatedly, the method which leads from the one-particle problem to the many-body problem in the case of Bose statistics' \citep[p.\ 297; our emphasis]{Jordan 1973}.
\end{quotation}
Since this was a conference honoring Dirac, other speakers can be forgiven for declaring Dirac to be the founding father of quantum field theory. Rudolf \citet{Peierls 1973} set the tone in his talk on the development of quantum field theory: ``When the principles of quantum mechanics were established it was clear that, for consistency, the electromagnetic radiation had to be described by a formalism based on the same principles. On  this subject, as on so many others, the first step was taken by Dirac" (p.\ 370).
The next day, Julian \citet{Schwinger 1973} followed suit in his report on quantum electrodynamics: ``Essentially everything I shall discuss today has its origin in one or another work of Dirac. That will not astonish you. It is unnecessary, of course, but nevertheless I remind you that {\it quantum} electrodynamics began in the famous paper of 1927.\footnote{At this point an editorial comment is inserted: 
``[A]ppropriately, this is the first paper in the collection, {\it Selected Papers on Quantum Electrodynamics} [Schwinger, 1958]."} Here Dirac first extended the methods of quantum mechanics to the electromagnetic field" (p.\ 414). Gregor Wentzel chaired this session and the conference proceedings also contain a reprint of  his review of quantum field theory for the Pauli memorial volume, which prominently mentions the {\it Dreim\"annerarbeit} and lists the early papers of Jordan and his collaborators in its bibliography \citep[p.\ 49 and pp.\ 74--75]{Wentzel 1960}. Neither Wentzel nor Jordan said anything in the discussions following the talks by Peierls and Schwinger.
One wonders, however, whether these celebratory distortions of history induced Jordan to submit his reminiscences of the early years of quantum mechanics to the conference proceedings instead of the musings on the expansion of the earth to which he had treated his colleagues at the conference itself.

More recent histories of quantum field theory---especially \citep{Darrigol 1986} and, drawing on Darrigol's work, \citep{Miller 1994} and \citep{Schweber 1994}---do full justice to Jordan's contributions.\footnote{See also \citep{Ehlers 2007} and \citep{Schroer 2007} specifically on Jordan.} Jordan also gets his due in a sketch of the history of quantum field theory from a modern point of view in \citep[sec.\ 1.2, pp.\ 15--31]{Weinberg 1995}, which includes a discussion of the fluctuation calculations in the {\it Dreim\"annerarbeit}.\footnote{\label{milonni}It is difficult to gauge both how well-known and how well-understood these calculations have been in the physics community since their publication in 1926. One data point is provided by \citep{Milonni 1981, Milonni 1984}. In 1981, this author derived a formula for energy fluctuations in a box of black-body radiation ({\it not} a subvolume of this box) that has the form of Einstein's 1909 fluctuation formula. He calls Einstein's formula a ``seminal result" that ``is important as the first clear example of a wave-particle duality" \citep[p.\ 177; the author cites \citep{Pais 1979} in this context (cf.\ note \ref{pais})]{Milonni 1981}.  He interprets the two terms in his fluctuation formula ``in terms of the fundamental processes of spontaneous and stimulated emission, and absorption" and writes that ``[t]his interpretation seems obvious in retrospect but has not, to the author's knowledge, been discussed previously" (ibid.). He does not mention the {\it Dreim\"annerarbeit}. In a paper on the wave-particle duality three years later in a volume in honor of Louis de Broglie's 90th birthday, \citet[pp.\ 39--41]{Milonni 1984} does mention the fluctuation calculations in the {\it Dreim\"annerarbeit}, though he has clearly missed that these calculations, like Einstein's, pertain to a {\it subvolume} and seems to be under the impression that they are equivalent to the calculations in \citep{Milonni 1981}. He acknowledges that, when he wrote this 1981 paper, he ``was not aware that the Born-Heisenberg-Jordan paper contained a discussion of the fluctuation formula" \citep[p. 62, note 27]{Milonni 1984}. Neither were the editors and referees of {\it American Journal of Physics} it seems.}  Our paper focuses on Jordan's use of the idea of field quantization in his fluctuation calculations but to understand the negative reactions of his co-authors and contemporaries to these calculations it is important to keep in mind that Jordan was virtually alone at first in recognizing the need for the extension of quantum theory to fields.

\subsection{The 1926 Smekal interlude}

What is suppressed in Jordan's later recollections is that in the course of 1926 he himself started to have second thoughts about second quantization and that \citep{Dirac 1927} also seems to have been important in dispelling his own doubts.\footnote{This interlude is also discussed in \citep[pp.\ 222--225]{Darrigol 1986}.}  In April 1926, Adolf Smekal published a paper criticizing the fluctuation calculations in the {\it Dreim\"annerarbeit}. Smekal argued that, when calculating energy fluctuations in radiation, one should take into account the interaction with matter emitting and absorbing the radiation. Without such interaction, he insisted, the radiation would not reach its equilibrium black-body frequency distribution and would not be detectable so that {\it a fortiori} fluctuations in its energy would not be observable. With the first of these two objections, Smekal put his finger on a step that is missing both in H.\ A.\ Lorentz's (1916) derivation of the classical formula for the mean square energy fluctuation in black-body radiation, and in Jordan's derivation of its quantum counterpart in the simple model of a string. To derive a formula for {\it thermal} fluctuations, one needs to consider a thermal ensemble of states. Both Lorentz and Jordan, however, only considered individual states and fail to make the transition to an ensemble of states. A clear indication of the incompleteness of their derivations is that the temperature does not appear anywhere. Smekal is quite right to insist that we consider the system, be it black-body radiation or oscillations in a string, in contact with an external heat bath. This does not mean, however, that the interaction with matter needs to be analyzed in any detail.  We can calculate the thermal fluctuations simply assuming that the system has somehow thermalized through interaction with matter. It should also be emphasized that the fluctuations in a small subvolume that Lorentz and Jordan are interested in do not come from the exchange of energy between radiation and matter but from radiation energy entering and leaving the subvolume. Smekal's second objection---that interaction with matter is needed to detect energy fluctuations in radiation---seems to have gained considerable traction with the authors of the {\it Dreim\"annerarbeit}, as one would expect given their Machian-positivist leanings.

In response to Smekal's criticism of their paper, the authors seem to have retreated to the position that their calculation was certainly valid for sound waves in a solid\footnote{In his lecture on specific heats at the first Solvay conference in 1911, \citet[p.\ 342]{Einstein 1914} had already made it clear that these fluctuation considerations also apply to solids \citep[p.\ 180]{Bach 1989}.} and that it was an open question for now whether it also applied to electromagnetic radiation. This is clear from a paper by \citet[p.\ 501, note 2]{Heisenberg 1926b} on fluctuation phenomena in crystal lattices and from a letter he simultaneously sent to Born, Jordan, and Smekal. As he told these three correspondents:
\begin{quotation}
Our treatment [i.e., in the {\it Dreim\"annerarbeit}] of fluctuation phenomena is undoubtedly applicable to the crystal lattice \ldots 
The question whether this computation of fluctuations can also be applied to a radiation cavity can, as Mr. Smekal emphasizes, not be decided at the moment, as a quantum mechanics of electrodynamical processes has not been found yet. Because of the formal analogy between the two problems (crystal lattice--cavity) I am personally inclined to believe in this applicability, but for now this is just a matter of taste.\footnote{Heisenberg to Born, Jordan, and Smekal, October 29, 1926 (AHQP).} 
\end{quotation}
A more definite stance would have to await the quantum-mechanical treatment of a full interacting system of radiation and matter. Dirac's paper provided such a treatment.

The retreat triggered by \citep{Smekal 1926} and the renewed advance after \citep{Dirac 1927} left some traces in Jordan's writings of this period.
Immediately after the discussion of his fluctuation considerations in his semi-popular history of the light-quantum hypothesis, we read:
\begin{quotation}
For light itself one can look upon the following thesis as the fundamental result of the investigation of Born, Heisenberg, and Jordan, namely that (as demanded by Pauli) a new {\it field concept} must be developed in which one applies the concepts of quantum mechanics to the oscillating field. But this thesis has in a sense shared the fate of the [fluctuation] considerations by Einstein, the elucidation of which served as its justification: for a long time---even among proponents of quantum mechanics---one either suspended judgement or rejected the thesis. It was accepted only when Dirac showed a year later that Einstein's [1917] laws for emission and absorption for atoms in a radiation field also follow necessarily and exactly from this picture [of quantized fields] \citep[pp.\ 195--196]{Jordan 1928}.
\end{quotation}
In a footnote appended to the next-to-last sentence, Jordan acknowledges that ``this general rejection" of field quantization had caused him to doubt it himself ``for a while" and that these doubts had found their way into his two-part overview of recent developments in quantum mechanics \citep{Jordan 1927a, Jordan 1927b}. In his presentation of his fluctuation considerations in the second part, \citet[pp.\ 642--643]{Jordan 1927b} accepts Heisenberg's criticism\footnote{See the second of the three passages quoted above from Jordan to Van der Waerden, April 10, 1962 (see note \ref{letter041062a}).} that it is unclear whether quantum mechanics as it stands applies to systems with an infinite number of degrees of freedom and, again following Heisenberg's lead, retreats to the claim that the analysis certainly holds for a lattice with a finite number of particles:
\begin{quotation}
Since a crystal is a system of  particles, nothing prevents us from applying the laws of quantum mechanics to it. Born, Heisenberg, and Jordan calculated the fluctuation of the oscillatory energy in a crystal lattice and arrived at the result that the matrix-theoretical description of these oscillations really does lead to the value of the mean square fluctuation required by thermodynamics \citep[p.\ 643]{Jordan 1927b}.
\end{quotation}
In the {\it Dreim\"annerarbeit}, the authors still confidently asserted that the same considerations that apply to a finite crystal lattice ``also apply if we go over to the limiting case of a system with infinitely many degrees of freedom and for instance consider the vibrations of an elastic body idealized to a continuum or finally of an electromagnetic cavity" (3M, p.\ 375). In a note added in proof to \citep{Jordan 1927b}, inserted right after the paragraph with the passage quoted above,  Jordan announces with obvious relief that Dirac's forthcoming paper completely vindicates the original generalization from a crystal lattice to electromagnetic radiation.

\subsection{Jordan's result as evidence for matrix mechanics}

The ambivalence of Born and Heisenberg about Jordan's fluctuation considerations, whatever its sources, is reflected in the use that is made of Jordan's result in the {\it Dreim\"annerarbeit}. Rather than hailing it as a seminal breakthrough in its own right, the authors make it subordinate to the overall aim of promoting matrix mechanics. As they announce at the end of the introduction, the recovery of Einstein's fluctuation formula ``may well be regarded as significant evidence in favour of the quantum mechanics put forward here" (3M, p.\ 325). After presenting the result, they comment: 
\begin{quotation}
If one bears in mind that the question considered here is actually somewhat remote from the problems whose investigation led to the growth of quantum mechanics, the result \ldots can be regarded as particularly encouraging for the further development of the theory (3M, p.\ 385). 
\end{quotation}
The one other accomplishment the authors explicitly identify as providing ``a strong argument in favour of the theory" is their derivation of the Kramers dispersion formula, ``otherwise obtained only on the basis of correspondence considerations" (3M, p.\ 333).  Since the new theory grew directly out of these correspondence considerations \citep{Duncan and Janssen}, it is not terribly surprising that it correctly reproduces this formula. The recovery of the Einstein fluctuation formula, which played no role in the construction of the theory, constitutes much more striking evidence.

Moreover, as the authors themselves emphasize, the way in which the theory reproduces the fluctuation formula is a particularly instructive illustration of the basic idea of {\it Umdeutung}. Before going into the details of the calculations, the authors already express the hope ``that the modified kinematics which forms an inherent feature of theory proposed here would yield the correct value for the interference fluctuations" (3M, p.\ 377). In the next-to-last paragraph of the paper, they make sure the reader appreciates that this hope has now been fulfilled:
\begin{quotation}
The reasons leading to the appearance [in the formula for the mean square energy fluctuation] of a term which is not provided by the classical theory are obviously closely connected with the reasons for [the] occurrence of a zero-point energy. The basic difference between the theory proposed here and that used hitherto in both instances lies in
the characteristic kinematics and not in a disparity of the mechanical laws. One could
indeed perceive one of the most evident examples of the difference between quantum-theoretical
kinematics and that existing hitherto on examining [the quantum fluctuation
formula], which actually involves no mechanical principles whatsoever (3M, p.\ 385).
\end{quotation}
With the exception of the final clause, which is best set aside as a rhetorical flourish, the authors' point is well taken. In the spirit of Heisenberg's groundbreaking paper, ``Quantum-theoretical re-interpretation of kinematic and mechanical relations," the fluctuation formula, the Kramers dispersion formula, and other results are obtained not through a change of the dynamical laws (the $q$'s and $p$'s for the oscillators representing the modes of the field satisfy the usual laws of Newtonian mechanics) but through a change of the kinematics (the nature of the $q$'s and $p$'s is changed).  In this particular case, this means that the wave equation for the string---the analogue of Maxwell's equations for this simple model---is taken over intact from the classical theory but that the displacement of the continuous string from its equilibrium state and the time derivatives of that displacement are no longer given by an infinite set of numbers but by an infinite set of infinite-dimensional matrices. 

As we shall see in our detailed reconstruction of Jordan's argument in sec.\ 3, the occurrence of a particle-like term in the formula for the mean square energy fluctuation once the classical system of waves in a string is subjected to the {\it Umdeutung} procedure is a direct consequence of  the zero-point energy found when the harmonic oscillator is quantized with the help of this procedure, a result already derived in the {\it Umdeutung} paper itself (Heisenberg, 1925, pp.\ 271--272; see also Born and Jordan, 1925, pp.\ 297--300). In his brief comments on the {\it Dreim\"annerarbeit} in a classic paper on Einstein and quantum physics, Stachel correctly identifies the zero-point energy as the key element in Jordan's recovery of Einstein's fluctuation formula. However, he does not mention that the zero-point energy itself is traced to the central new feature of the new theory, the {\it Umdeutung} of position and momentum as matrices subject to a quantum commutation relation.\footnote{\citet[p.\ 379]{Stachel 1986} writes: ``By introduction of the so-called zero-point energy of the proper vibrations into which the field was analyzed, Jordan---who wrote this section of the paper---claimed to show that the new matrix mechanics, when applied to fields, could produce not only the quadratic (classical wave) term but also the linear (classical particle) term."} Without this additional piece of information, it looks as if Jordan obtained his result simply by sleight of hand. \citet[p.\ 212]{Kojevnikov 1990} does mention that the zero-point energy is itself a consequence of the new theory, though the point could have done with a little more emphasis. \citet[p.\ 222]{Darrigol 1986},  in his brief characterization of Jordan's calculation, stresses the role of non-commutativity  and does not explicitly mention the zero-point energy at all.\footnote{\label{abelian} \citet[p.\ 199]{Bach 1989} acknowledges that ``sometimes it is pointed out that the cause of the occurrence of the two terms [in the fluctuation formula] lies in the non-commutativity of the observables of the quantum theory," but claims that this is mistaken since the observables relevant to the fluctuation problem supposedly form an Abelian subalgebra (ibid., pp.\ 199, 202). This claim is simply false. The relevant observables, the operator for the energy of the whole system and the operator for the energy in part of the system in a narrow frequency range, do not commute (see sec.\ 3.2, the discussion following eq.\ (\ref{q28})).}

In 1926, in the letter to Born, Jordan, and Smekal from which we already quoted above, Heisenberg made it clear that the fluctuation calculations were important to him only insofar as they provided evidence for matrix mechanics (especially for the basic idea of {\it Umdeutung}) and, by this time, against wave mechanics:
\begin{quotation}
For the crystal lattice the quantum-mechanical treatment [of fluctuations] undoubtedly means essential progress. This progress does not lie therein that one has found the mean square fluctuation; that one already had earlier and is obvious on the basis of general thermodynamical considerations if one introduces quantum jumps. The progress, in fact, lies therein that quantum mechanics allows for the calculation of these fluctuations without explicit consideration of quantum jumps on the basis of relations between $q$, $q^\prime$ etc. This amounts to a strong argument for the claim that quantum-mechanical matrices are the appropriate means for representing discontinuities such as quantum jumps (something that does not become equally clear in the wave-mechanical way of writing things). The calculation of our {\it Dreim\"annerarbeit} thus provided an element of support for the correctness of quantum mechanics.\footnote{Heisenberg to Born, Jordan, and Smekal, October 29, 1926 (AHQP).}
\end{quotation} 
Now that the calculation had served that purpose, Heisenberg clearly preferred to leave it behind and move on.\footnote{In the same letter, Heisenberg claimed that he only reluctantly agreed to the publication of the section on fluctuation phenomena of the {\it Dreim\"annerarbeit}:
``I wanted to give up on publishing our  {\it Dreim\"annernote}, because all polemics are abhorrent to me in the bottom of my soul [{\it weil mir jede Polemik im Grund meiner Seele v\"ollig zuwider ist}] and because I no longer saw any point worth fighting for." \citet{Darrigol 1986} renders the  italicized clause as ``because any polemics cut me to the quick" adding that  ``this is a rhetorical figure" (p.\ 223).} In fact, he closes his letter reminding his colleagues that ``there are so many beautiful things in quantum theory at the moment that it would be a shame if no consensus could be reached on a detail like this" ({\it eine solche Einzelheit}; ibid.). 

\subsection{Jordan on the intrinsic value of his result}

For Jordan, the intrinsic value of the fluctuation result was obviously much higher than for Heisenberg. Still, Jordan also had a tendency to make his result subservient to a larger cause, albeit quantum field theory rather than matrix mechanics. We already quoted the passage from his history of the light-quantum hypothesis in which he wrote that the fundamental importance of the result was that it brought out the need for a quantum theory of fields \citep[pp.\ 195--196]{Jordan 1928}. In this same article, however, as in various subsequent publications, Jordan also did full justice to the intrinsic importance of the result as having resolved Einstein's conundrum of the wave-particle duality of light. In his history of the light-quantum hypothesis, he wrote: 
\begin{quotation}
[I]t turned out to be superfluous to explicitly adopt the light-quantum hypothesis: We explicitly stuck to the wave theory of light and only changed the kinematics of cavity waves quantum-mechanically. From this, however, the characteristic light-quantum effects emerged automatically as a consequence \ldots This is a whole new turn in the light-quantum problem. It is not necessary to include the picture [{\it Vorstellung}] of light quanta among the assumptions of the theory. One can---and this seems to be the natural way to proceed---start from the wave picture. If one formulates this with the concepts of quantum mechanics, then the characteristic light-quantum effects emerge as necessary consequences from the general laws [{\it Gesetzm\"a\ss igkeiten}] of quantum theory \citep[p.\ 195]{Jordan 1928}.
\end{quotation}
As Jordan undoubtedly realized, ``the characteristic light-quantum effects" referred to in this passage do not include those that involve interaction between the electromagnetic field and matter, such as the photoelectric effect or the Compton effect. His argument only applies to free radiation. As such, however, it does explain why the Einstein fluctuation formula contains both a particle and a wave term and why light quanta are subject to Bose's odd new statistics. That the latter also speaks in favor of Jordan's approach is explicitly mentioned in the {\it Dreim\"annerarbeit} (3M, pp.\ 376--379), in \citep[p.\ 182]{Jordan 1928}, and, in more detail, in \citep[p.\ 220]{Jordan 1936}, (an attempt at) an elementary textbook on quantum mechanics.\footnote{For discussion of the connection to Bose statistics, see \citep[p.\ 221]{Darrigol 1986}. Darrigol also quotes from a letter from Jordan to Erwin Schr\"odinger that can be dated to the summer of 1927, in which Jordan briefly reiterates this point (ibid., p.\ 224).} The connection with Bose statistics is not mentioned in \citep{Jordan 1929}, his contribution to the proceedings of the Charkow conference on unified field theory and quantum mechanics. That is probably just because the paper is not about quantum statistics but about quantum field theory. As such it provides a prime example of Jordan using his fluctuation result as a means to an end (i.e., the promotion of quantum field theory), but it also contains a particularly crisp statement of the intrinsic value of the result complete with an uncharacteristically immodest  assessment of its momentous character: 
\begin{quotation}
Einstein drew the conclusion that the wave theory would necessarily have to be replaced or at least supplemented by the corpuscular picture. With our findings [{\it Feststellungen}], however, the problem has taken a completely different turn. We see that it is not necessary after all to abandon or restrict the wave theory in favor of other models [{\it Modellvorstellungen}]; instead it just comes down to reformulating [{\it \"ubertragen}] the wave theory in quantum mechanics. The fluctuation effects, which prove the presence of corpuscular light quanta in the radiation field, then arise automatically as consequences of the wave theory.  The old and famous problem how one can understand waves and particles in radiation in a unified manner can thus in principle be considered as taken care of [{\it erledigt}] \citep[p.\ 702]{Jordan 1929} 
\end{quotation}
The strong confidence conveyed by that last sentence probably reflects that with \citep{Dirac 1927} the tide had decisively turned for Jordan's pet project of quantizing fields. What militates against this conclusion, at least {\it prima facie}, is that Jordan's language in \citep[pp.\ 398--399]{Born and Jordan 1930} is more subdued again, but that could be because he feared he would not get a  more exuberant statement past his teacher and co-author. As is clear in hindsight (see sec.\ 2.2), Born was not paying much attention and Jordan need not have worried, but Jordan probably did not know that in 1930. In any event, the language of the Charkow proceedings, including the triumphant last sentence,  is recycled {\it verbatim} in \citep[p.\ 222]{Jordan 1936}. 

None of these texts, of course, ever drew anywhere near the attention the {\it Dreim\"annerarbeit} drew. Reading them {\it seriatim}, moreover, one can appreciate the complaint by \citet{Pauli 1930}  to the effect that Jordan was starting to sound like a broken record (see sec.\ 2.2). Jordan did himself a disservice not only by agreeing to present his result merely as a piece of evidence for matrix mechanics in the {\it Dreim\"annerarbeit}, but also by trying to make up for that mistake too many times. That Jordan   routinely pressed the result into service in his promotion of quantum field theory may also have hurt the recognition of its significance in its own right. 

\subsection{Einstein's reaction to Jordan's result}

Even if the intrinsic merit of his result had been brought out more clearly in his paper with Born and Heisenberg, Jordan would still not have convinced Einstein, who, as the person responsible for the riddle, should have been especially interested in the solution Jordan claimed to have found. That the recovery of the fluctuation formula was explicitly presented as evidence for matrix mechanics, however, almost certainly made matters worse. His co-authors may have had reservations about applying the {\it Umdeutung} procedure to fields; Einstein did not want to apply it to {\it anything}. As Jordan told Van der Waerden: ``Einstein was really the only physicist from whom I could expect the acknowledgment [{\it Feststellung}] that with this result a big problem in physics had really been brought to its solution. But, although he reacted very friendly and kindly, Einstein on his part was disinclined to consider matrix mechanics as trustworthy."\footnote{Jordan to Van der Waerden, December 1, 1961 (AHQP). As the reference to Einstein's friendliness suggests, politics did not play a role in Einstein's negative reaction.} He told Kuhn the same thing: ``One might have imagined that Einstein would have been pleased [with Jordan's result] but Einstein's attitude toward matrix mechanics was that he was having none of it"
(AHQP interview with Jordan, session 3, p.\ 9).

Rather than dismissing Jordan's result out of hand because of its origin in Heisenberg's ``large quantum egg,"\footnote{Einstein to Ehrenfest, September 30, 1925, quoted, for instance, in \citep[p.\ 566]{Foelsing 1997}.} Einstein had a very specific objection \citep[p.\ 222]{Darrigol 1986}. 
He had been given proofs of the {\it Dreim\"annerarbeit} and about a week after the paper was published, he wrote to Ehrenfest: 
\begin{quotation}
There cannot be a zero-point energy of black-body radiation. I consider the argument by Heisenberg, Born, and Jordan (fluctuations) pertaining to this to be fallacious, for one thing because the probability of large fluctuations (for example, for finding the total energy in the subvolume $U$ of [the total volume] $V$) does certainly not come out right.\footnote{Einstein to Ehrenfest, February 12, 1926, quoted in \citep[p.\ 212]{Kojevnikov 1990}. Kojevnikov translates ``nicht richtig herauskommt" as ``cannot be derived."} 
\end{quotation}
The example of a large fluctuation mentioned in parentheses is the one \citet{Einstein 1905} used in the paper in which he first introduced the light-quantum hypothesis. To show that  black-body radiation in the Wien regime of high frequencies behaves like an ideal gas of non-interacting point particles, he derived a formula for the phenomenally small probability that all the energy of black-body radiation in some container would spontaneously be confined to a small subvolume of that container. \citet[pp.\ 161--162]{Jordan 1928} called this formula ``Einstein's first fluctuation law" to distinguish it from the fluctuation formula we have been discussing so far, which he called ``Einstein's second fluctuation law." 

In early April, Einstein sent Jordan a postcard with the same objection that he had run by Ehrenfest:
\begin{quotation}
The thing with the fluctuations is fishy [{\it faul}]. One can indeed calculate the average magnitude of fluctuations  with the zero-point term $\frac{1}{2} h \nu$, but not the probability of a very large fluctuation. For weak (Wien) radiation, the probability, for instance, that all radiation [in some narrow frequency range around $\nu$] is found in a subvolume $V$ of the total volume $V_0$ is
$
W = \left( V/V_0 \right)^{E/h \nu}.
$
This can evidently not be explained with the zero-point term although the expression is secure [{\it gesichert}] on thermodynamical grounds.\footnote{Einstein to Jordan, April 6, 1926 (AHQP), quoted, for instance, in \citep[Vol.\ 3, p.\ 156]{Mehra Rechenberg} in a slightly different translation.}
\end{quotation}
In a postscript Einstein added somewhat disingenuously: ``Other than that, however, I am greatly impressed with matrix theory." Presumably, Einstein's problem with the zero-point energy was that, if the energy of each quantum of frequency $\nu$ were $\frac{3}{2}h\nu$ rather than $h\nu$, the exponent in the expression for $W$ above can no longer be interpreted as the number of light quanta $N$ and black-body radiation in the Wien regime would no longer behave as an ideal gas of $N$ particles. The zero-point energy, however, does not come into play here, since the energy $E$ in the exponent of Einstein's formula is a thermal average of the excitation energy, the difference between the full energy and the zero-point energy, which the authors of the  {\it Dreim\"annerarbeit} misleadingly call the ``thermal energy" (3M, p.\ 377, p.\ 384).\footnote{Cf.\ the discussion following Eq.\ (\ref{q21b}) in sec.\ 3.2.} 

What about Einstein's more general objection that Jordan could not calculate the probability of an arbitrary fluctuation? In his history of the light-quantum hypothesis, \citet[p.\ 194]{Jordan 1928} conceded that this was indeed practically impossible in the theory {\it as it stood}. Of course, it is no condemnation of Jordan's derivation of the second fluctuation law that he did not derive the first as well. \citet[p.\ 194, p.\ 201]{Jordan 1928} claimed to have a better line of defense. In a paper published the year before \citep[pp.\ 772--774]{Jordan 1927c}, he had further developed the theory and outlined how one could now calculate the probability of arbitrary fluctuations.\footnote{In \citep[p.\ 399]{Born and Jordan 1930}, Jordan promises that it will be shown in Part II of the book that a quantum-mechanical wave theory also reproduces Einstein's first fluctuation law correctly. Unfortunately, Part II never appeared. Since \citep[p.\ 399]{Born and Jordan 1930} was itself the sequel to \citep{Born 1925}, \citet{Pauli 1930} sarcastically began his review by pointing out that ``[t]his book is the second volume in a series in which goal and purpose of the $n$th volume is always made clear through the virtual existence of the $(n+1)$th volume." This review helped ensure that the ``$n=3$"-volume never saw the light of day.} As far as we know, Einstein's response to this development, if there ever was any, has not been preserved. 

Einstein never accepted Jordan's result and maintained to the end of his life that the puzzle of the wave-particle duality of light remained without a solution. As he told his old friend Michele Besso a few years before he died:
\begin{quotation}
All these fifty years of conscious brooding have brought me no closer to the answer to the question ``What are light quanta?" Nowadays every Tom, Dick, and Harry [{\it jeder Lump}] thinks he knows it, but he is mistaken.\footnote{Einstein to Besso, December 12, 1951, quoted, for instance, in \citep[p.\ 133, p.\ 138]{Klein 1979}.}
\end{quotation}
Einstein scholars typically quote such pronouncements approvingly in recounting the story of the light-quantum hypothesis and the wave-particle duality of light.\footnote{See, e.g., \citep[pp.\ 379--380]{Stachel 1986}.} Their message, it seems, is that, as wrong as Einstein turned out to be about other aspects of quantum mechanics, he was {\it right} about the wave-particle duality of light. In our estimation, he was just being stubborn. Quantum electrodynamics provides a perfectly satisfactory solution to Einstein's 1909 riddle of the wave-particle duality of light. Jordan was the first to hit upon that solution. The problem of the infinite zero-point energy, to be sure, is still with us in the guise of the problem of the vacuum energy, but that is a different issue. Recall, moreover, that Jordan avoided the problem of infinite zero-point energy altogether by deriving the mean square energy fluctuation in a finite frequency range. 

\subsection{Why Jordan's result has not become more famous}

In the course of our analysis in secs.\ 2.1--2.7, we identified several factors that help explain why Jordan's derivation of Einstein's fluctuation formula has not become part of the standard story of wave-particle duality. Before moving on to the actual calculations, we collect these factors here. First, there is the cloud of suspicion that has always surrounded the result. Then there is the tendency, most notably in the case of Heisenberg but also in the case of Jordan himself, to downplay the intrinsic value of the result as resolving the conundrum of the wave-particle duality of light and to present it instead as an argument for either matrix mechanics or the need for a quantum theory of fields. This is reflected in the historical literature where Jordan's result has meanwhile found its proper place in histories of quantum field theory but is hardly ever mentioned in histories of the light-quantum hypothesis and the wave-particle duality of light.

\section{Reconstruction of and commentary on Jordan's derivation of Einstein's fluctuation formula}

Jordan borrowed a simple model from \citet[pp.\ 367--373]{Ehrenfest 1925a} to analyze the problem of energy fluctuations in black-body radiation. He considered a string of length $l$ fixed at both ends.  This can be seen as a one-dimensional analogue of an electromagnetic field forced to vanish at the conducting sides of a box. The wave equation for the string---the analogue of the free Maxwell equations for this simple model---is:
\begin{equation}
\frac{\partial^2 u}{\partial t^2} - \frac{\partial^2 u}{\partial x^2} = 0,
\end{equation}
where $u(x, t)$ is the displacement of the string at position $x$ and time $t$ and where the velocity of propagation is set equal to unity. The boundary conditions $u(0,t) = u(l, t) = 0$ for all times $t$ express that the string is fixed at both ends.
The general solution of this problem can be written as a Fourier series (3M, Ch.\ 4, Eqs.\ (41) and ($41'$)):
\begin{equation}
u(x,t) = \sum_{k=1}^\infty q_k(t) \sin{(\omega_k x)},
\label{2}
\end{equation}
with angular frequencies
\begin{equation}
\omega_k \equiv  \frac{k \pi}{l},
\label{3a}
\end{equation}
and  Fourier coefficients (3M, Ch.\ 4, Eq.\ (44)) 
\begin{equation}
q_k(t) = a_k \cos{(\omega_k t + \varphi_k)}.
\label{3}
\end{equation}
The Hamiltonian for the string is (3M, Ch.\ 4, Eq.\ (42) [$u^2$ should be $\dot{u}^2$]):
\begin{equation}
H = \frac{1}{2} \int_0^l dx \left( \dot{u}^2 + u_x^2 \right), 
\label{1}
\end{equation}
The dot indicates a time derivative and the subscript $x$ a partial derivative with respect to $x$. The terms $\dot{u}^2$ and $u_x^2$ are the analogues of the densities of the electric and the magnetic field, respectively, in this simple model of  black-body radiation. 
The Hamiltonian in Eq.\ (\ref{1}) can be transformed into an infinite sum of Hamiltonians $H_j$ for uncoupled individual harmonic oscillators of mass $l/2$ (3M, Ch.\ 4, Eq.\ (42)). Inserting Eq.\ (\ref{2}) for $u(x,t)$ in Eq.\ (\ref{1}),
we find (3M, Ch.\ 4, Eq.\ (41)):
\begin{eqnarray}
\lefteqn{ H  =  \frac{1}{2} \int_0^l dx \sum_{j,k =1}^\infty \left(  \dot{q}_j(t) \dot{q}_k(t)
\sin{(\omega_j x)} \sin{(\omega_k x)}  \right.} \nonumber \\
 & & \;\;\;\;\;\;\;\;\;\;\;\;\;\;\;\;\;\;\;\;\;\;\;\;\;\;  \left.
 + \;\;  \omega_j \omega_k q_j(t) q_k(t) \cos{(\omega_j x)} \cos{(\omega_k x)} \right).
\label{3b}
\end{eqnarray}
The functions $\{ \sin{(\omega_k x)} \}_k$ in  Eq.\ (\ref{2}) are orthogonal on the interval $(0, l)$, i.e.,
\begin{equation}
\int_0^l dx   \sin{(\omega_j x)}  \sin{(\omega_k x)}  =  \frac{l}{2} \delta_{jk}.
\label{3c}
\end{equation}
The same is true for the functions $\{ \cos{(\omega_k x)} \}_k$. It follows that the integral in Eq.\ (\ref{3b}) will only give contributions for $j = k$ (as can be verified explicitly by substituting $l$ for $a$ in Eq.\ (\ref{11}) below). The double sum thus turns into the single sum:
\begin{equation}
H  = \sum_{j=1}^\infty \frac{l}{4} \left(  \dot{q}_j^2(t) + \omega_j^2 q_j^2(t)  \right) = \sum_{j=1}^\infty H_j.
\label{4}
\end{equation}
The vibrating string can thus be replaced by an infinite number of uncoupled oscillators, one for every mode of the string, with characteristic angular frequencies $\omega_j = j (\pi/l)$ (see Eq.\ (\ref{3a})). This shows that the distribution of the energy over all the frequencies of these oscillators is constant in time. Since there is no coupling between the oscillators, there is no mechanism for transferring energy from one mode to another. The spatial distribution of the energy in a given frequency range over the length of the string, however, varies in time. We study the fluctuations of the energy in a narrow frequency range in a small segment of the string. The total energy in that frequency range will be constant but the fraction located in that small segment will fluctuate. Jordan derived an expression for the mean square energy fluctuation of this energy, first in classical theory, then in matrix mechanics.

\subsection{Classical calculation}

Changing the upper boundary  of the integral in Eq.\ (\ref{3b}) from $l$ to $a$ ($a \ll l$) and restricting the sums over $j$ to correspond to a narrow angular frequency range $(\omega, \omega + \Delta \omega)$, we find the instantaneous energy in that frequency range in a small segment $(0, a) \subset (0, l)$ of the string. This quantity is simply called $E$ in the paper (cf.\ 3M, Ch.\ 4, Eq.\ (43)). We add the subscript $(a, \omega)$:  
\begin{eqnarray}
\lefteqn{ E_{(a, \omega)}(t) = \frac{1}{2} \int_0^a dx \sum_{j,k}  \left(  \dot{q}_j(t) \dot{q}_k(t) 
\sin{(\omega_j x)} \sin{(\omega_k x)} \right.  } \nonumber \\
 & & \;\;\;\;\;\;\;\;\;  \;\;\;\;\;\;\;\;\; \;\;\;\;\;\;\;\;\; \;\;\;\;\;\;\;\;\;  \;\;\;\;\;\;\;\;\; \left.
 + \;\;  \omega_j \omega_k q_j(t) q_k(t) \cos{(\omega_j x)} \cos{(\omega_k x)} \right),
\label{5}
\end{eqnarray} 
where the sums over $j$ and $k$ are restricted to the finite range of integers satisfying $\omega < j (\pi/l) < \omega+\Delta\omega$ and $ \omega < k (\pi/l) < \omega+\Delta\omega$. Unless we explicitly say that sums run from 1 to $\infty$, {\it all} sums in what follows are restricted to this finite range. This restriction also appears to be in force in many summations in this section of the {\it Dreim\"annerarbeit} even though they are all {\it written} as infinite sums. The sums in Eqs. (43), (45), ($46'$), ($46''$), and (47) in the paper (3M, pp.\ 381--382) should all be over these finite rather than over an infinite range of integers.

There are several clear indications in this section of the paper that the authors are in fact considering a small frequency range. The clearest statement is their description of the situation with black-body radiation which the string is supposed to represent:
\begin{quotation}
If there is communication between a volume $V$ and a very large volume such that waves which have {\it frequencies which lie within a small range $\nu$ to $\nu + d\nu$} can pass unhindered from one to the other, whereas for all other waves the volumes remain detached, and if $E$ be the energy of the waves with frequency $\nu$ in $V$, then according to Einstein the mean square deviation \ldots
can be calculated (3M, p.\ 379, our emphasis).
\end{quotation}
Two pages later, in Eq.\ (43) the same symbol $E$ is used for what we more explicitly write as $E_{(a, \omega)}$ and immediately below this equation it says in parentheses:  ``under the explicit assumption that all wavelengths {\it which come into consideration} are small with respect to $a$" (3M, p.\ 381, our emphasis). 

The functions $\{ \sin{(\omega_k x)} \}_k$ and the functions $\{ \cos{(\omega_k x)} \}_k$ are not orthogonal on the interval $(0,a)$, so both terms with $j=k$ and terms with $j \neq k$ will contribute to the instantaneous energy $E_{(a, \omega)}(t)$ in Eq.\ (\ref{5}). First consider the  $(j=k)$ terms. On the assumption that $a$ is large enough for the integrals over $\sin^2{(\omega_j x)}$ and $ \cos^2{(\omega_j x)}$ to be over many periods corresponding to $\omega_j$, these terms are given by:
\begin{equation}
E_{(a, \omega)}^{(j=k)}(t) = 
\frac{a}{4} \sum_{j}  \left( \dot{q}_j^2(t) + \omega_j^2  q_j^2(t) \right) =  \frac{a}{l} \sum_{j} H_j(t).
\label{6aa}
\end{equation}
Since we are dealing with a system of uncoupled oscillators, the energy of the individual oscillators is constant. Since all terms $H_j(t)$ are constant, $E_{(a, \omega)}^{(j=k)}(t)$ is constant too and equal to its time average:\footnote{A bar over any quantity denotes the time average of that quantity. The argument that follows, leading to Eqs.\ (\ref{6a}) and (\ref{9}) can also be made in terms of averages over the phases $\varphi_k$ in the Fourier coefficients in Eq.\ (\ref{3}). }
\begin{equation}
E_{(a, \omega)}^{(j=k)}(t) = \overline{ E_{(a, \omega)}^{(j=k)}(t) }
\label{6a}
\end{equation}
Since the time averages $\overline{\dot{q}_j(t) \dot{q}_k(t) }$ and $\overline{q_j(t) q_k(t)}$ vanish for $j \neq k$, the ($j \neq k$) terms in Eq.\ (\ref{5}) do not contribute to its time average:
\begin{equation}
\overline{ E_{(a, \omega)}^{(j \neq k)}(t) } = 0.
\label{6ac}
\end{equation}
The time average of Eq.\ (\ref{5}) is thus given by the $(j=k)$ terms:
\begin{equation}
E_{(a, \omega)}^{(j=k)}(t) = \overline{E_{(a, \omega)}(t)}.
\label{6}
\end{equation}
Combining Eqs.\ (\ref{6aa}) and (\ref{6}), we see that the time average of the energy in the frequency range $(\omega, \omega + \Delta \omega)$ in the small segment $(0,a)$ of the string is just the fraction $(a/l)$ of the (constant) total amount of energy in this frequency range  in the entire string.

From Eq.\  (\ref{6}) it follows that the $(j \neq k)$ terms in Eq.\ (\ref{5}) give the instantaneous deviation $\Delta E_{(a, \omega)}(t)$ of the energy in this frequency range in the segment $(0,a)$ of the string from its mean (time average) value:
\begin{equation}
\Delta E_{(a, \omega)}(t) \equiv E_{(a, \omega)}(t) -  \overline{E_{(a, \omega)}(t)} = E_{(a, \omega)}^{(j \neq k)}(t).
\label{9}
\end{equation}
With the help of the trigonometric relations $\sin{\alpha}\sin{\beta} = \frac{1}{2} (\cos{(\alpha - \beta)} - \cos{(\alpha + \beta)})$ and  $\cos{\alpha}\cos{\beta} = \frac{1}{2} (\cos{(\alpha - \beta)} + \cos{(\alpha + \beta)})$, we integrate the $(j \neq k)$ terms  in Eq.\ (\ref{5}) to find $\Delta E_{(a, \omega)}$. From now on, we suppress the time dependence of $\Delta E_{(a, \omega)}$, $q_j$ and $\dot{q}_j$.
\begin{eqnarray}
\Delta E_{(a, \omega)} & = & \frac{1}{4}  \int_0^a dx \sum_{j \neq k} \left( \dot{q}_j \dot{q}_k 
\left[ \cos{((\omega_j - \omega_k)x)} - \cos{((\omega_j + \omega_k)x)}\right]  \right. \nonumber \\
& & \;\;\;\;\;\;\;\;\;\;\;\;\;\; \left. +  \;  \omega_j \omega_k q_j q_k 
\left[ \cos{((\omega_j - \omega_k)x)} + \cos{((\omega_j + \omega_k)x)}\right] \right) \nonumber \\
 & & \label{11} \\
 & = & \frac{1}{4}  \sum_{j \neq k} \left( \dot{q}_j \dot{q}_k 
 \left[ \frac{\sin{((\omega_j - \omega_k)a)}}{\omega_j - \omega_k} -  
 \frac{\sin{((\omega_j + \omega_k)a)}}{\omega_j + \omega_k} \right] \right. \nonumber \\  
& & \;\;\;\;\;\;\;\;\;\;\;\; \left. + \;  \omega_j \omega_k q_j q_k 
\left[ \frac{\sin{((\omega_j - \omega_k)a)}}{\omega_j - \omega_k} +  
 \frac{\sin{((\omega_j + \omega_k)a)}}{\omega_j + \omega_k} \right]  \right). \nonumber
\end{eqnarray}
Defining the expressions within square brackets as (cf.\ 3M, Ch.\ 4, Eq.\ (45$^\prime$))
\begin{eqnarray}
K_{jk} & \equiv &  \frac{\sin{((\omega_j - \omega_k)a)}}{\omega_j - \omega_k} -  
 \frac{\sin{((\omega_j + \omega_k)a)}}{\omega_j + \omega_k}, \nonumber \\
 & & \label{11a} \\
K^\prime_{jk} & \equiv & \frac{\sin{((\omega_j - \omega_k)a)}}{\omega_j - \omega_k} +  
 \frac{\sin{((\omega_j + \omega_k)a)}}{\omega_j + \omega_k}, \nonumber
\end{eqnarray}
we can write this as (cf.\ 3M, Ch.\ 4, Eq.\ (45)):
\begin{equation}
\Delta E_{(a, \omega)} =  \frac{1}{4}  \sum_{j \neq k} \left( \dot{q}_j \dot{q}_k K_{jk}
+ \omega_j \omega_k q_j q_k K^\prime_{jk} \right).
\label{12}
\end{equation}
Note that both $K_{jk}$ and $K^\prime_{jk}$ are symmetric: $K_{jk} = K_{kj}$ and $K^\prime_{jk} = K^\prime_{kj}$.
We now compute the mean square fluctuation of the energy in the segment $(0, a)$  in the frequency range $(\omega, \omega + \Delta \omega)$. Denoting the two parts of the sum in Eq.\ (\ref{12}) as $\Delta E_{1_{(a, \omega)}}$ and $\Delta E_{2_{(a, \omega)}}$, respectively, we find (3M, Ch.\ 4, Eq.\ (46)):
\begin{equation}
\overline{\Delta E_{(a, \omega)}^2}  =  \overline{\Delta E_{1_{(a, \omega)}}^2} + \overline{\Delta E_{2_{(a, \omega)}}^2} + 
\overline{\Delta E_{1_{(a, \omega)}} \Delta E_{2_{(a, \omega)}}} + \overline{\Delta E_{2_{(a, \omega)}} \Delta E_{1_{(a, \omega)}}}.
\label{13}
\end{equation}
Classically, the last two terms are obviously equal to one another. In quantum mechanics we have to be more careful. So it is with malice of forethought that we wrote these last two terms separately. The first two terms are given by (3M, Ch.\ 4, Eq.\ (46$'$))
\begin{eqnarray}
\overline{\Delta E_{1_{(a, \omega)}}^2} + \overline{\Delta E_{2_{(a, \omega)}}^2} & = &
\frac{1}{16}  \sum_{j \neq k}  \sum_{j' \neq k'} \left(
\overline{ \dot{q}_j \dot{q}_k  \dot{q}_{j'} \dot{q}_{k'}} K_{jk} K_{j'k'}  \right. \nonumber \\
 & & \;\;\;\;\;\;\;\;\;\;\;\;\;\;\;\;\; + \;\; \left.
\overline{q_j q_k q_{j'} q_{k'}}  \omega_j \omega_k \omega_{j'} \omega_{k'}
 K'_{jk} K'_{j'k'} \right);
\label{14}
\end{eqnarray}
the last two by (3M, Ch.\ 4, Eq.\ (46$''$))
\begin{eqnarray}
\lefteqn{ \overline{\Delta E_{1_{(a, \omega)}} \Delta E_{2_{(a, \omega)}}} + \overline{\Delta E_{2_{(a, \omega)}} \Delta E_{1_{(a, \omega)}}} = }
 & & \nonumber \\
 & &\label{15} \\
 & & \;\;\;\;\;\;\;\;\;\; \frac{1}{16}  \sum_{j \neq k}  \sum_{j' \neq k'} 
\left( \overline{\dot{q}_j \dot{q}_k q_{j'} q_{k'} } \omega_{j'} \omega_{k'}  K_{jk} K'_{j'k'}
 +
 \overline{q_j q_k \dot{q}_{j'} \dot{q}_{k'}} \omega_j \omega_k K'_{jk} K_{j'k'}\right).  \nonumber
\end{eqnarray}
Since $q_k(t) = a_k \cos{(\omega_k t + \varphi_k)}$ and $\omega_k = k (\pi/l)$ (see Eqs.\ (\ref{3a}) and (\ref{3})), it would seem that the time averages of the products of four $q$'s or four $\dot{q}$'s in Eq.\ (\ref{14}) vanish unless $(j=j', k=k')$ or $(j=k', k=j')$. This is not strictly true. Writing $\cos{\omega_k}t = \frac{1}{2} \left( e^{i \omega_k t} + e^{- i \omega_k t} \right)$, we see that there will in principle be non-vanishing contributions whenever $\pm \,\, j \pm k \pm j' \pm k' = 0$.\footnote{This problem does not arise if we consider phase averages instead of time averages. Since the phases $\varphi_j$, $\varphi_k$, $\varphi_{j'}$, and $\varphi_{k'}$ in the $q$'s and $\dot{q}$'s are statistically independent, the only contributions to Eqs.\ (\ref{14}) and (\ref{15}) with time averages replaced by phase averages
come from terms in the quadruple sum over $(j \neq k, j' \neq k')$ with either $(j=j', k=k')$ or $(j=k', k=j')$.}  In the real physical situation, however, the $\omega$'s will not {\it exactly} be integral number times $(\pi/l)$. This effectively precludes the possibility of satisfying the identity through index combinations for which $j \neq k \neq j' \neq k'$. Hence, the only index combinations contributing in the real physical situation are indeed just $(j=j', k=k')$ and $(j=k', k=j')$. They both give the same contribution. Hence (3M, Ch.\ 4, Eq.\ (47)):
\begin{equation}
\overline{\Delta E_{1_{(a, \omega)}}^2} + \overline{\Delta E_{2_{(a, \omega)}}^2} =
\frac{1}{8}  \sum_{j \neq k} \left(  \overline{\dot{q}_j^2} \, \overline{\dot{q}_k^2} K_{jk}^2
+ \overline{q_j^2} \, \overline{q_k^2} \omega_j^2 \omega_k^2 K_{jk}^{\prime 2} \right),
\label{16}
\end{equation}
where we used that, for $j \neq k$, averages of products such as $\overline{ \dot{q}_j^2 \dot{q}_k^2 }$ are products of the averages
$\overline{\dot{q}_j^2}$ and $\overline{\dot{q}_k^2}$. The time averages in Eq.\ (\ref{15}), with two $q$'s and two $\dot{q}$'s rather than four $q$'s or four $\dot{q}$'s, vanish even if $(j=j', k=k')$ or $(j=k', k=j')$. These index combinations produce time averages of expressions of the form $\sin{(\omega_jt + \varphi_j)} \cos{(\omega_jt + \varphi_j)} = \frac{1}{2} \sin{(2(\omega_jt + \varphi_j))}$ and these vanish. So, in the classical theory, Eq.\ (\ref{16}) gives the total mean square fluctuation.

To evaluate the mean square averages of the $q$'s and $\dot{q}$'s in Eq.\ (\ref{16}), we use the virial theorem, which says that the time average of the kinetic energy of any one of the oscillators in Eq.\ (\ref{4}) is equal to the time average of its potential energy:
\begin{equation}
\frac{l}{4} \overline{ \dot{q}_j^2(t) } = \frac{l}{4} \omega_j^2 \overline {q_j^2(t)} = \frac{1}{2} H_j.
\label{16b}
\end{equation}
It follows that
\begin{equation}
\overline{\Delta E_{(a, \omega)}^2}  =
 \frac{1}{2l^2} \sum_{j \neq k} H_j H_k \left( K^2_{jk} + K^{\prime \, 2}_{jk} \right).
\label{q14}
\end{equation}
We now assume 
that the energies $H_j$ of the oscillators of characteristic frequency $\omega_j$ vary smoothly with $j$. This assumption, which is not made explicit in the {\it Dreim\"annerarbeit}, does not hold for arbitrary distributions of the total energy over the various frequencies, but it does hold for a broad class of them.
As long as $H_j$ varies smoothly with $j$, we can replace the double sum over $j$ and $k$ by a double integral over the continuous variables $\omega$ and $\omega'$. Since $\omega_j = (\pi/l) j$ (see Eq.\ (\ref{3a})), a sum over $j$ turns into $(l/\pi)$ times an integral over $\omega$. Next, we introduce the continuous counterparts $K_{\omega \omega'}$ and $K'_{\omega \omega'}$ of $K_{jk}$ and $K'_{jk}$. They are obtained by changing $\omega_j$ and $\omega_k$ in Eq.\ (\ref{11a}) into $\omega$ and $\omega'$.
When we integrate over the square of $K_{\omega \omega'}$, the contribution coming from the square of second term, which has $\omega + \omega'$ in the denominator, is negligibly small compared to the contribution coming from the square of the first term, which has $\omega - \omega'$ in the denominator (the integral over the product of the first and the second term vanishes). The same is true for integrals over the square of $K'_{\omega \omega'}$.\footnote{It can be shown that neglecting the terms with $(\omega_j + \omega_k)$ in the denominator in these integrals causes a relative error of order $\displaystyle{ \frac{\Delta\omega}{\omega}\, \frac{1}{a\omega}}$, a product of two factors much smaller than 1.\label{small}} In such integrals $K^2_{\omega \omega'}$ and $K^{\prime \, 2}_{\omega \omega'}$ can thus be replaced by the squares of their (identical) first terms. Moreover, if $a$ is very large compared to the wavelengths associated with the frequencies in the narrow range 
$(\omega, \omega + \Delta \omega)$, we can set (cf.\ 3M, Ch.\ 4, Eq.\ (48)):
\begin{equation}
\int d \omega' f(\omega') \frac{ \sin^2{((\omega - \omega')a)}}{(\omega - \omega')^2} = \int d\omega' f(\omega') \pi a \delta(\omega - \omega') = \pi a f(\omega),
\label{17}
\end{equation}
where $\delta(x)$ is the Dirac delta function and $f(x)$ is an arbitrary smooth function. 
When we make all these substitutions, Eq.\ (\ref{q14}) turns into:
\begin{equation}
\overline{\Delta E_{(a, \omega)}^2} = \frac{1}{2l^2} \int d\omega \int d\omega' \left( \frac{l}{\pi} \right)^2
2 \pi a \delta(\omega - \omega') H_\omega H_{\omega'} = \frac{a}{\pi} \int d\omega H^2_\omega,
\label{18}
\end{equation}
where the integrals are over the interval $(\omega, \omega + \Delta \omega)$.\footnote{The corresponding integrals in Eqs.\ ($47'$), (49), and (50) are from 0 to $\infty$, just as the sums in Eqs.\ (43), (45), ($46'$), ($46''$), and (47). After Eq.\ (49), equivalent to our Eq.\ (\ref{18}), and Eq.\ (50) for what in our notation would be $\overline{E_a}$, the authors write:  ``In order to obtain [the thermodynamical mean square energy fluctuation and the mean energy] we have merely to {\it extract those parts referring to $d \nu = d \omega/2 \pi$}" (3M, p.\ 383, our emphasis). This is another clear indication that the authors intended to compute the mean square energy fluctuation in a narrow frequency range.} Instead of integrating we can multiply the integrand by $\Delta \omega = 2 \pi \Delta \nu$. If, in addition, we replace functions of $\omega$ by functions of $\nu$, we can write Eq.\ (\ref{18}) as:
\begin{equation}
\overline{\Delta E_{(a, \nu)}^2} = 2a \Delta \nu H^2_\nu.
\label{18a}
\end{equation}
Finally, we replace $H_\nu$ by the average energy in the frequency range $(\nu, \nu + \Delta \nu)$ in the segment $(0,a)$ of the string, using the relation
\begin{equation}
\overline{E_{(a, \nu)}} = N_\nu \left( \frac{a}{l} H_\nu \right),
\label{19}
\end{equation}
where $N_\nu$ is the number of modes in the interval $(\nu, \nu + \Delta \nu)$. Eq.\ (\ref{3a}) tells us that $\pi N_\nu/l = 2 \pi \Delta \nu$, so
\begin{equation}
N_\nu = 2l \Delta \nu.
\label{20}
\end{equation}
It follows that 
\begin{equation}
H_\nu = \frac{\overline{ E_{(a, \nu)} }}{2 a \Delta \nu}.
\label{21}
\end{equation}
Inserting this into Eq.\ (\ref{18a}), we arrive at
\begin{equation}
\overline{ \Delta E_{(a, \nu)}^2 } = 
\frac{ \overline{E_{(a, \nu)} }^2 }{2a \Delta \nu},
\label{22}
\end{equation}
for the mean square energy fluctuation in the small segment $(0, a)$ of the string  in the narrow frequency range $(\nu, \nu + \Delta \nu)$. This is the analogue of the formula for the mean square energy fluctuation in a narrow frequency range in a small part of a larger volume containing black-body radiation.  Eq.\ (\ref{22}) shows that the mean square fluctuation in the energy is proportional to the mean energy squared. 

Eq.\ (\ref{22}) holds for any state of the string in which there is a smooth distribution of the total energy over the various modes. However, Eq.\ (\ref{22}) is {\it not} the formula for the {\it thermal} mean square energy fluctuation, the quantity that should be compared to Einstein's fluctuation formula of 1909. A clear indication of this is that the temperature $T$ does not appear anywhere in its derivation. What we need is not a formula for $\overline{ \Delta E_{(a, \nu)}^2 }$ in an individual state but a formula for the average $\langle \overline{ \Delta E_{(a, \nu)}^2 } \rangle$ in a thermal ensemble of states. Without this extra step, the derivation is incomplete. The authors of the {\it Dreim\"annerarbeit} did not take this extra step, neither in the classical nor in the quantum-mechanical version of the calculation. In their defense, when \citet{Lorentz 1916}  derived his formula for the mean square fluctuation of the energy in a small subvolume of a box with classical electromagnetic radiation, he only derived the  analogue of Eq.\ (\ref{22}) for that system and did not calculate the average of this quantity in a thermal ensemble of states either. In Lorentz's calculation one also looks in vain for the temperature.\footnote{That a fluctuation formula can be derived without reference to thermodynamical considerations was noted by Max von \citet[p.\ 199]{Laue 1915c} \citep[p.\ 199]{Bach 1989}. This observation also plays an important role in Smekal's (1926, p.\ 321) criticism of this section of the {\it Dreim\"annerarbeit} (cf.\ sec.\ 2.4).}

Unlike the authors of the {\it Dreim\"annerarbeit}, we shall calculate the thermal average of the mean square energy fluctuation formulae they derived, both for the classical formula  (\ref{22})  and, at the end of sec.\ 3.2, for the quantum formula (\ref{q28}). Classically, a state of the string is fully specified by the amplitudes $a_k$ and phases $\varphi_k$ of the infinite number of modes of the string. The thermal average of any observable $O(a_1, a_2, \ldots \varphi_1, \varphi_2, \ldots)$ of the system, which will be some function of these amplitudes and phases, is given by the average over a canonical ensemble of such states:\footnote{Given that the amplitudes and the phases are continuous quantities, the sums in this equation are symbolic representations of integrals with a measure determined by the transformation from $\{ q_i, p_i \}$ to $\{ a_i,  \varphi_i\}$.}
\begin{equation}
\langle O(a_1, a_2, \ldots \varphi_1, \varphi_2, \ldots)  \rangle
= \frac{\displaystyle{\sum_{\{ a_1, a_2, \ldots \varphi_1, \varphi_2, \ldots\}} O(\ldots) e^{- \displaystyle{\beta E_{\{ a_1, a_2, \ldots \varphi_1, \varphi_2, \ldots\}}}}}}{
\displaystyle{\sum_{\{ a_1, a_2, \ldots \varphi_1, \varphi_2, \ldots\}} e^{- \displaystyle{\beta E_{\{ a_1, a_2, \ldots \varphi_1, \varphi_2, \ldots\}}}}}},
\label{23}
\end{equation} 
with $\beta \equiv 1/kT$, where $k$ is Boltzmann's constant and $T$ is the temperature. The underlying physical picture is that we imagine the string to be coupled to an infinite heat bath at temperature $T$. We compute the ensemble average of the expression for $\overline{\Delta E_{(a, \omega)}^2}$ in Eq.\ (\ref{q14}). The only part that we need to be careful about is the product $H_j H_k$. So we set $O$ in Eq.\ (\ref{23}) equal to:
\begin{equation}
O(a_1, a_2, \ldots \varphi_1, \varphi_2, \ldots) = H_j (a_j, \varphi_j) H_k (a_k, \varphi_k)
\label{24}
\end{equation}
The energy of the string in a given state is just the sum of the Hamiltonians for all the different modes in that state:
\begin{equation}
E_{\{ a_1, a_2, \ldots \varphi_1, \varphi_2, \ldots\}} = \sum_{i=1}^\infty H_i(a_i, \varphi_i).
\label{25}
\end{equation} 
It follows that the denominator in Eq.\ (\ref{23}) can be rewritten as:
\begin{eqnarray}
\displaystyle{\sum_{\{ a_1, a_2, \ldots \varphi_1, \varphi_2, \ldots\}} e^{- \displaystyle{\beta E_{\{ a_1, a_2, \ldots \varphi_1, \varphi_2, \ldots\}}}}} & = & 
\displaystyle{\sum_{\{ a_1, a_2, \ldots \varphi_1, \varphi_2, \ldots\}} e^{- \displaystyle{\beta \sum_{i=1}^\infty H_i(a_i, \varphi_i)}}}
\nonumber \\
 & & \label{26} \\
 & = & \prod_{i=1}^\infty \left( \sum_{\{ a_i, \varphi_i \}}
e^{ \displaystyle{ -\beta H_i(a_i, \varphi_i) } } \right).\nonumber
\end{eqnarray}
For all but the $j^{th}$ and the $k^{\rm th}$ mode, the $i^{\rm th}$ factor in the denominator of Eq.\ (\ref{23}) with $O=H_j H_k$ cancels against an identical factor in the numerator. Eq.\ (\ref{23}) thus reduces to the product of two factors of the exact same form, one for the $j^{\rm th}$ and one for the $k^{\rm th}$ mode. The $j^{th}$ mode gives:
\begin{equation}
\frac{\displaystyle{\sum_{\{ a_j, \varphi_j\}} H_j(a_j, \varphi_j) e^{- \displaystyle{\beta H_j(a_j, \varphi_j)}}}}{\displaystyle{\sum_{\{ a_j, \varphi_j\}} e^{- \displaystyle{\beta H_j(a_j, \varphi_j)}}}}
\label{27}
\end{equation}
This is just the ensemble average $\langle H_j \rangle$ of the $j^{\rm th}$ mode. The same is true for the $k^{\rm th}$ mode. The equipartition theorem tells us that the average energy of a one-dimensional simple harmonic oscillator at temperature $T$ is equal to $kT$. The modes of the string thus satisfy the analogue of the classical Rayleigh-Jeans law for black-body radiation. Using that $\langle H_j H_k \rangle = \langle H_j  \rangle \langle  H_k \rangle$, we see that the ensemble average of Eq.\ (\ref{q14}) is given by:
\begin{equation}
\langle \overline{\Delta E_{(a, \omega)}^2} \rangle =
 \frac{1}{2l^2} \sum_{j \neq k} \langle H_j \rangle \langle H_k \rangle \left( K^2_{jk} + K^{\prime \, 2}_{jk} \right).
\label{29}
\end{equation}
Repeating the steps that got us from Eq.\ (\ref{q14}) to Eq.\ (\ref{22}), we arrive at:
\begin{equation}
\langle \overline{ \Delta E_{(a, \nu)}^2 } \rangle = 
\frac{ \langle \overline{E_{(a, \nu)} } \rangle^2 }{2a \Delta \nu},
\label{30}
\end{equation}
This is the classical formula for the {\it thermal} mean square fluctuation of the energy in a narrow frequency range in a small segment  of the string. Note that in this case the assumption we need to make to replace sums by integrals is that $\langle H_i \rangle$ varies smoothly with $i$, which will certainly be true. In fact,  it is a constant: $\langle H_i \rangle = kT$. At first sight, it may be surprising that Eq.\ (\ref{30}) for the thermal average of the mean square energy fluctuation has the same form as Eq.\ (\ref{22}) for the mean square energy average in an individual state. The reason for this can be gleaned from Eq.\ (\ref{q14}). The entire
contribution to the mean square energy fluctuation comes from off-diagonal (i.e., $j \neq k$) terms involving the product of two distinct, and therefore thermally uncorrelated, modes. This is true for the set of uncoupled oscillators that replaces the string. It is not true for arbitrary systems.


\subsection{Quantum-mechanical calculation}

In the {\it Dreim\"annerarbeit} (3M, pp.\ 383--384), the classical calculation covered in sec.\ 3.1 is translated into a quantum-mechanical one with the help of Heisenberg's {\it Umdeutung} procedure. The $q$'s and the $\dot{q}$'s thus become matrices that do not always commute. The zero-point energy of the harmonic oscillator is a direct consequence of this feature. Another consequence is that the terms in Eq.\ (\ref{15}), which vanished in the classical case, do contribute to the mean square energy fluctuation in the quantum case. Both this contribution and the zero-point energy of the modes of the string, it turns out, are essential for correctly reproducing the analogue of the particle term in Einstein's fluctuation formula in the simple model used in the {\it Dreim\"annerarbeit}. 

As we mentioned in the introduction, the key point of Heisenberg's {\it Umdeutung} paper is that the new quantities representing position and momentum in the new theory still satisfy the classical equations of motion. So the solution for the harmonic oscillator is just the solution of the classical equation of motion for the harmonic oscillator, $\ddot{q}_k(t) = - \omega_k^2 q_k(t)$, but now reinterpreted as an equation for matrices. This solution is given by \citep[p.\ 139, Eq.\ (530)]{Baym 1969}:\footnote{Setting $q_k(0) = a_k \cos{\varphi_k}$ and $p_k(0) = - (l \omega_k a_k/2) \sin{\varphi_k}$ in Eq.\ (\ref{q4}), and 
interpreting $q_k(t)$, $\dot{q}_k(t)$, $q_k(0)$, and $p_k(0)$ as ordinary numbers, we recover Eq.\ (\ref{3}):
$$
q_k(t)  = a_k \left( \cos{\varphi_k}  \cos{ \omega_k t}  -  \sin{\varphi_k} \sin{ \omega_k t}  \right) = a_k \cos{(\omega_k t + \varphi_k)}.
$$
In the quantum case, we no longer have the freedom to choose arbitrary phases $\varphi_k$ that we had in the classical case. Accordingly, we can no longer average over such phases. In the {\it Dreim\"annerarbeit} phase averages are simply {\it defined} as the diagonal part of the quantum-theoretical matrix for the relevant quantity in a basis of energy eigenstates (3M, p.\ 383).}
\begin{equation}
q_k(t)  =  q_k(0) \cos{ \omega_k t}  +  \frac{2p_k(0)}{l \omega_k}  \sin{ \omega_k t}.
\label{q4}
\end{equation}
Differentiating this equation, we find:
\begin{equation}
\dot{q}_k(t)  =  \frac{2p_k(0)}{l}  \cos{ \omega_k t}  -  \omega_k q_k(0) \sin{ \omega_k t}.
\label{q4a}
\end{equation}
In these equations, $q_k(t)$, $\dot{q}_k(t)$, $q_k(0)$, and $p_k(0)$ are all matrices, satisfying the canonical equal-time commutation relations $[q_j(t), q_k(t)] = 0$, $[p_j(t), p_k(t)] = 0$, and $[q_j(t), p_k(t)] = i \hbar \delta_{jk}$. 

An energy eigenstate of the system is given by specifying the values of the infinite set $\{n_k\}$ of excitation levels  of all the modes of the string. The total energy $E$ of the system in the state $\{n_k\}$ is the expectation value of the Hamilton operator $H$ for the whole system in that state. This is the diagonal matrix element $H(\{n_k\}, \{n_k\})$ (which in modern notation would be $\langle n_k | H | n_k \rangle$):
\begin{equation}
H(\{n_k\}, \{n_k\}) = \sum_{k=1}^\infty \left( n_k + \frac{1}{2} \right)  \hbar \omega_k.
\label{q6}
\end{equation}
The zero-point energy in Eq.\ (\ref{q6}) is clearly infinite. However, as long as we continue to restrict ourselves to a narrow frequency range, the contribution to the zero-point energy will be perfectly finite.

To find such quantities as the mean energy and the mean square energy fluctuation in a small part of the string and in a narrow frequency range, we first retrace our steps in the classical calculation given above, keeping in mind that $q$'s, $p$'s, and $\dot{q}$'s are no longer numbers but---in modern terminology---operators. We then evaluate the expectation values of the resulting operators in an energy eigenstate of the full system, specified by the excitation levels $\{n_k\}$. As in Eq.\ (\ref{q6}), these expectation values are the diagonal matrix elements of the operators  in a basis of energy eigenstates. In the {\it Dreim\"annerarbeit}  the argument is formulated entirely in terms of such matrix elements, but  it becomes more transparent if we phrase it in terms of operators and their expectation values. In Ch.\ 3 of the {\it Dreim\"annerarbeit}, on ``the connection with the theory of eigenvalues of Hermitian forms,"  the authors get close to the notion of operators acting on a state space but they do not use it in the more physical sections of the paper. They clearly recognized, however, that the matrix elements they computed are for states specified by excitation levels of the infinite set of  oscillators. The final step, which is not in the {\it Dreim\"annerarbeit}, is to compute the average of the quantum expectation value of the relevant operator in a canonical ensemble of energy eigenstates.

Most of the intermediate results in the classical calculation can be taken over unchanged with the understanding that we are now dealing with operators rather than numbers. Replacing the $q$'s and $\dot{q}$'s (or, equivalently, the $p$'s) in Eq.\ (\ref{5}) for $E_{(a, \omega)}$ by the corresponding operators and renaming the quantity  $H_{(a, \omega)}$, we find the Hamilton operator for the small segment $(0, a)$ of the string in the narrow angular frequency range $(\omega, \omega + \Delta \omega)$. This is a perfectly good Hermitian operator, which corresponds, at least in principle, to an observable quantity. We want to emphasize that this is true despite the restriction to a narrow frequency range. 

Our first goal is to find the operator $\overline{\Delta H_{(a, \omega)}^2}$ for the mean square fluctuation of $H_{(a, \omega)}$. Eqs.\ (\ref{6}) and (\ref{9}) for the $(j=k)$ terms and the $(j \neq k)$ terms in $E_{(a, \omega)}$, respectively, remain valid for the $(j=k)$ terms and the $(j \neq k)$ terms of $H_{(a, \omega)}$. The $(j=k)$ terms give the operator for the time average of the energy in the segment $(0, a)$  in the frequency range $(\omega, \omega + \Delta \omega)$:
\begin{equation}
H_{(a, \omega)}^{(j=k)} = \overline{H_{(a, \omega)}}.
\label{q6a}
\end{equation}
The $(j \neq k)$ terms  give the operator for the instantaneous energy fluctuation in this segment and in this frequency range:
\begin{equation}
H_{(a, \omega)}^{(j \neq k)} = \Delta H_{(a, \omega)} =  H_{(a, \omega)} - \overline{H_{(a, \omega)}}.
\label{q7}
\end{equation}
This quantity is still given, {\it mutatis mutandis}, by Eq.\ (\ref{12}). As before, we split it into two parts, $\Delta H_{(a, \omega)} = \Delta H_{1_{(a, \omega)}} + \Delta H_{2_{(a, \omega)}}$. Thus, as in Eq.\ (\ref{13}), the operator $\overline{\Delta H_{(a, \omega)}^2}$ for the mean square fluctuation of the energy in the small segment $(0, a)$ in the frequency range $(\omega, \omega + \Delta \omega)$ is given by four terms.
The first two terms are still given by Eq.\ (\ref{q14}),\footnote{As the authors explicitly note, the virial theorem, which was used to get from Eq.\ (\ref{16}) to Eq.\ (\ref{q14}),  remains valid in matrix mechanics (3M, pp.\  343 and 383).} the last two by Eq.\ (\ref{15}) (with $E$ replaced by $H$). The latter vanished in the classical case but not in the quantum case (3M, p.\ 384).\footnote{\label{incomprehensible}This is the step that \citet[p.\ 263]{Born and Fuchs 1939a} complained involved ``quite incomprehensible reasoning" (cf.\ note \ref{born complaint}). They wrote: ``The error in the paper of Born, Heisenberg, and Jordan is in the evaluation of the terms $\Delta_1 \Delta_2 + \Delta_2 \Delta_1$ (see formula (46$''$) [3M, p.\ 382; our Eq.\ (\ref{15})]). On [3M, p.\ 382] it is correctly stated that in the classical calculation the mean value of this quantity over all phases vanishes. This is also true in the quantum mechanical calculation as is apparent from formula (46$''$). [On the bottom half of 3M, p.\ 384], however, $\overline{\Delta_1 \Delta_2 + \Delta_2 \Delta_1}$ reappears again with a non-vanishing value [cf.\ our Eq.\ (\ref{q18})] and it is shown that it gives rise to an additional term by means of quite incomprehensible reasoning. It is just this term which transforms the correct formula (2.1) [the mean square energy fluctuation for classical waves; cf.\ our Eq.\ (\ref{22})] into the thermodynamical formula (1.6) [Einstein's fluctuation formula]. But from the standpoint of wave theory this formula (1.6) is certainly wrong" (ibid.). As we shall see, there is nothing wrong with this step in the argument in the {\it Dreim\"annerarbeit}. We suspect that what tripped up Born in 1939 was the distinction between phase averages and time averages in the {\it Dreim\"annerarbeit}.}  These terms now give identical contributions for the index combinations $(j=j', k=k')$ and $(j=k', k=j')$ with $j \neq k$ and $j' \neq k'$. The quadruple sum in Eq.\ (\ref{15}) reduces to the double sum:
\begin{eqnarray}
\lefteqn{ \overline{\Delta H_{1_{(a, \omega)}} \Delta H_{2_{(a, \omega)}}} + \overline{\Delta H_{2_{(a, \omega)}} \Delta H_{1_{(a, \omega)}}} }
& & \nonumber \\
 \label{7a} \\ 
 & & \;\;\;\;\;\;\;\ =  \frac{1}{8} \sum_{ j \neq k} \left(  
\overline{ \dot{q}_j q_j } \; \overline{ \dot{q}_k q_k}
 + \overline{ q_j \dot{q}_j } \; \overline{q_k \dot{q}_k}
 \right) \omega_j \omega_k K_{jk} K'_{jk},
\nonumber
\end{eqnarray}
where we used that $q_j$ commutes with $\dot{q}_k$ as long as $j \neq k$. 
The two terms in Eq.\ (\ref{7a}), it turns out, give identical contributions. We focus on the first. We compute the time average $\overline{\dot{q}_j q_j}$. Using Eqs.\ (\ref{q4}) and (\ref{q4a}), we find that
$$
\overline{\dot{q}_j q_j} = \overline{\left(  \frac{2p_j(0)}{l}  \cos{ \omega_j t}  -  \omega_j q_j(0) \sin{ \omega_j t} \right) 
\left(  q_j(0) \cos{ \omega_j t}  +  \frac{2p_j(0)}{l \omega_j}  \sin{ \omega_j t} \right)},
$$
which reduces to
\begin{equation}
\overline{\dot{q}_j q_j} = \frac{1}{l} \left( p_j(0) q_j(0) - q_j(0) p_j(0)\right).
\label{q16}
\end{equation}
Classically, $p$ and $q$ commute, but in quantum theory we have $[q_j(0), p_j(0)] = i \hbar$, so that
\begin{equation}
\overline{\dot{q}_j q_j} = - \frac{i \hbar}{l}.
\label{q17}
\end{equation}
The time average $\overline{ q_j \dot{q}_j}$ is likewise given by $i \hbar/l$. Inserting these results into Eq.\ (\ref{7a}), we find
\begin{equation}
\overline{\Delta H_{1_{(a, \omega)}} \Delta H_{2_{(a, \omega)}}} + \overline{\Delta H_{2_{(a, \omega)}} \Delta H_{1_{(a, \omega)}}}
= - \frac{\hbar^2}{4l^2} \sum_{j \neq k} \omega_j \omega_k K_{jk} K'_{jk}.
\label{q18}
\end{equation}
When we add the contributions to $\overline{\Delta H_{(a, \omega)}^2}$ coming from Eq.\ (\ref{q18}) to those coming from Eq.\ (\ref{q14}), we find
\begin{equation}
\overline{\Delta H_{(a, \omega)}^2} =
\frac{1}{l^2}  \sum_{j \neq k} \left(  H_j H_k \frac{1}{2} \left( K^2_{jk} + K^{' \, 2}_{jk}\right) - 
\frac{\hbar^2}{4} \omega_j \omega_k K_{jk} K'_{jk} \right).
 \label{q21}
\end{equation}
Replacing both $ \frac{1}{2} \left( K^2_{jk} + K^{' \, 2}_{jk} \right)$ and $K_{jk} K'_{jk}$ by $\sin^2{((\omega_j - \omega_k)a)}/(\omega_j - \omega_k)^2$
(cf.\ the paragraph before Eq.\ (\ref{17})), we can rewrite Eq.\ (\ref{q21}) as 
\begin{equation}
\overline{\Delta H_{(a, \omega)}^2} =
\frac{1}{l^2}  \sum_{j \neq k} \left(  H_j H_k  - 
\frac{\hbar^2}{4} \omega_j \omega_k  \right)   \frac{\sin^2{((\omega_j - \omega_k)a)}}{(\omega_j - \omega_k)^2}.
 \label{q30}
 \end{equation}
The next step---and the final step in the {\it Dreim\"annerarbeit}---is to evaluate the expectation value of the operator $\overline{\Delta H_{(a, \omega)}^2}$ in the state $\{n_i\}$. This is the diagonal matrix element, $\overline{\Delta H_{(a, \nu)}^2}(\{n_i\}, \{n_i\})$.\footnote{We remind the reader that the authors of the {\it Dreim\"annerarbeit} do not explicitly distinguish between operators and their expectation values. This is a source of possible confusion at this point. The authors write: ``we denote those parts of $\overline{\Delta^2}$ [rendered in bold] which belong to a given frequency $\nu$ as $\overline{\Delta^2}$ [not rendered in bold]" (3M, p.\ 384). Without any further information, one can read this either as a restriction (in our notation) of the operator $\overline{\Delta H_a^2}$ to the operator $\overline{\Delta H_{(a, \omega)}^2}$ or as a restriction of the states $\{ n_i \}$ in the matrix element $\overline{\Delta H_a^2}(\{n_i\}, \{n_i\})$ to states in which only modes in the frequency interval $\omega < i (\pi/l) < \omega+\Delta\omega$ are present (i.e., $n_i = 0$ for all frequencies $i (\pi/l)$ outside that narrow range). Since the latter reading makes no sense (we are interested in states  with excitations over the whole frequency spectrum), we assume that the former reading is what the authors had in mind. We are grateful to J\"urgen Ehlers for alerting us to this ambiguity.} 
Using that
\begin{equation}
H_j(\{ n_i \}, \{ n_i \}) = \left( n_j + \frac{1}{2} \right) \hbar \omega_j,
\label{q21c}
\end{equation}
we find that
\newpage
\begin{eqnarray}
\overline{\Delta E_{(a, \omega)}^2} & \equiv & \overline{\Delta H_{(a, \omega)}^2}(\{n_i\}, \{n_i \}) \nonumber \\
 & & \nonumber \\
& = &   \frac{1}{l^2}  \sum_{j \neq k} \left(  \left( n_j + \frac{1}{2} \right) \left( n_k + \frac{1}{2} \right)  - 
\frac{1}{4} \right) \hbar^2 \omega_j \omega_k     \frac{\sin^2{((\omega_j - \omega_k)a)}}{(\omega_j - \omega_k)^2}.  \nonumber \\
 & & \nonumber \\
 & = & \frac{1}{l^2}  \sum_{j \neq k} \left( n_j n_k + \frac{1}{2} \left( n_j + n_k \right)  \right) \hbar^2 \omega_j \omega_k   \frac{\sin^2{((\omega_j - \omega_k)a)}}{(\omega_j - \omega_k)^2}.
\label{q24} 
 \end{eqnarray}
 We thus see that the contribution to the mean square fluctuation coming from the second term in Eq.\ (\ref{q21}), which comes from the non-commutativity of $q$ and $p$, cancels the square of the zero-point energy in the contribution coming from  the first term.
 
Eq.\ (\ref{q24}) also illustrates the problem that \citet{Heisenberg 1931} drew attention to a few years later (see sec.\ 2.2). If we let $j$ and $k$ run from 1 to $\infty$ instead of restricting them to some finite interval, the double sum in Eq.\ (\ref{q24}) diverges. The problem comes from the terms with $(n_j + n_k)$; the contribution coming from the terms with $n_j n_k$ will still be perfectly finite, at least after we have made the transition from individual states to a thermal ensemble of states. In that case, the excitation level $n_i$ drops off exponentially with $i$ (see Eq.\ (\ref{q37})), so the double sum over the terms with $n_j n_k$ will quickly converge. This is not the case for the terms with just $n_j$ or just $n_k$. For a fixed value of $j$, for instance,\footnote{For fixed values of $k$, we run into the same problem.} the  double sum over the term with $n_j$ in Eq.\ (\ref{q24}) will reduce to the single sum:
\begin{equation}
\frac{n_j \hbar^2 \omega_j}{2l^2}   \sum_{k=1 \; (k \neq j)}^{\infty} \omega_k   \frac{\sin^2{((\omega_j - \omega_k)a)}}{(\omega_j - \omega_k)^2},
\label{q24a} 
\end{equation}
This sum is logarithmically divergent.\footnote{For large enough $k$, $\omega_j = j (\pi/l)$ will become negligible compared to $\omega_k = k (\pi/l)$ and the sum will be of the form 
$$
\sum_{k=1}^{\infty}  \frac{\sin^2{(k x/2)}}{k}= \sum_{k=1}^{\infty} \left( \frac{1}{2k} - \frac{\cos{(kx)}}{2k} \right)
$$
where $x \equiv 2 \pi a/l$. Using the standard formula
$$
\sum_{k=1}^{\infty} \frac{\cos{(kx)}}{k}= -\frac{1}{2} \ln{(2(1-\cos{(x)}))},
$$  
we see that the second term is finite. The first term, however, 
$
\sum_{k=1}^{\infty} 1/2k
$,
diverges logarithmically.} Following Heisenberg's suggestion in 1931, we can remedy this divergence if we replace the sharp edge of the segment of the string at $a$ by a  smooth edge. Integration over the segment $(0, a)$ of the string is equivalent to integration over the whole string if we multiply the integrand by the theta-function $\vartheta(a - x)$ (defined as: $\vartheta(\xi) = 0$ for $\xi < 0$ and $\vartheta(\xi) = 1$ for $\xi \geq 0$). The Fourier coefficients for this theta-function do not fall off fast enough if $j$ or $k$ go to infinity. This is why the factors $K_{jk}$ and $K'_{jk}$ in Eq.\ (\ref{11a}) do not fall off fast enough either if $j$ or $k$ go to infinity. If we replace the theta-function by a smooth, infinitely differentiable function, the problem disappears, since in that case the Fourier transform will fall off faster than any power of the transform variables $j$ or $k$. We emphasize that as long as the sums in Eq.\ (\ref{q24}) are restricted to a finite frequency interval, the result is finite without any such remedy.

As in the classical calculation (cf.\ Eqs.\ (\ref{q14})--(\ref{18a})), we  make the transition from sums to integrals. We can do this as long as the excitation levels $n_j$ vary smoothly with $j$. We can then replace the double sum over $j$ and $k$ by $(l/\pi)^2$ times a double integral over $\omega$ and $\omega'$ and $n_j$ and $n_k$ by $n_\omega$ and $n_{\omega'}$. We can also replace $\sin^2{((\omega - \omega')a)}/(\omega - \omega')^2$ by $\pi a \delta(\omega - \omega')$ (see Eq.\ (\ref{17})). Eq.\ (\ref{q24}) then turns into:
\begin{eqnarray}
\overline{\Delta E_{(a, \omega)}^2}
& = &
\frac{a}{\pi} \int d\omega \int d\omega' \; \delta(\omega - \omega')
 \left(  n_\omega n_{\omega'} + \frac{1}{2} \left( n_\omega  + n_{\omega'} \right) \right) \hbar^2 \omega \omega'  \nonumber \\
 & & \label{q21a} \\
 & = & \frac{a}{\pi} \int d\omega \left(  n_\omega^2 + n_\omega  \right) \hbar^2 \omega^2.  \nonumber
 \end{eqnarray}
Replacing integration over the interval $(\omega, \omega + \Delta \omega)$ by multiplication by $\Delta \omega = 2 \pi \Delta \nu$ and writing all quantities as functions of $\nu$ rather than $\omega$, we find:
\begin{equation}
\overline{\Delta E_{(a, \nu)}^2}  \equiv \overline{\Delta H_{(a, \nu)}^2}(\{ n_\nu \}, \{ n_\nu \}) = 2a \Delta \nu \left( (n_\nu h \nu)^2 +  (n_\nu h \nu) h \nu \right).
 \label{q21b}
\end{equation}
We now introduce the 
excitation energy, the difference between the total energy and the zero-point energy. Jordan and his co-authors call this  the ``thermal energy"  (3M, p.\ 377, p.\ 384). This terminology is misleading. It has nothing to do with temperature. The term `thermal energy' suggests  that the authors consider (what we would call) a thermal ensemble of energy eigenstates, while in fact they are only dealing with pure states. We therefore prefer the term `excitation energy'. The excitation energy $E_\nu$ in the narrow frequency range $(\nu, \nu + \Delta \nu)$ in the entire string in the state $\{ n_\nu \}$ is:
\begin{equation}
E_\nu = N_\nu (n_\nu  h \nu)  = 2 l \Delta \nu (n_\nu h \nu)
\end{equation}
where we used that $N_\nu = 2l \Delta \nu$ is the number of modes between $\nu$ and $\nu + \Delta \nu$ (see Eq.\ ({\ref{20})).
On average there will be a fraction $a/l$ of this energy in the small segment $(0,a)$ of the string (3M, p.\ 384, equation following Eq.\ (54)):\footnote{The time average $\overline{E_{(a, \nu)}}$ of the excitation energy  in the narrow frequency range $(\nu, \nu + \Delta \nu)$ in the small segment $(0,a)$ of the string in the state $\{ n_\nu \}$  is the expectation value of the operator $\overline{H_{(a, \nu)} - \frac{1}{2} h \nu}$ in that state.} 
\begin{equation}
\overline{E_{(a, \nu)}} = \frac{a}{l} E_\nu = 2 a \Delta \nu (n_\nu h \nu),
\label{q25}
\end{equation}
Substituting $\overline{E_{(a, \nu)}}/2a\Delta \nu$ for $n_\nu  h \nu$ in Eq.\ (\ref{q21b}), we arrive at the final result of this section of the {\it Dreim\"annerarbeit} (3M, Ch.\ 4, Eq.\ (55)):
\begin{eqnarray}
\overline{\Delta E_{(a, \nu)}^2} &
=  & 2a \Delta \nu \left( \left( \frac{\overline{E_{(a, \nu)}}}{2a\Delta \nu} \right)^2 
 + \left( \frac{\overline{E_{(a, \nu)}}}{2a\Delta \nu} \right) h \nu \right) \nonumber \\
 & &  \label{q28} \\
 & = &   \frac{\overline{E_{(a, \nu)}}^2}{2a\Delta \nu} + \overline{E_{(a, \nu)}} h \nu. \nonumber
\end{eqnarray}
Like Eq.\ (\ref{22}) in the classical case, Eq.\ (\ref{q28}) holds for any state with a smooth distribution of energy over frequency. Unlike the classical formula, however, Eq.\ (\ref{q28}) has exactly the same form as the fluctuation formula that Einstein derived on the basis of considerations in thermodynamics and statistical mechanics and Planck's law for the frequency distribution of black-body radiation. The first term has the form of the classical wave term (cf.\ Eq.\ (\ref{22})); the second term has the form of the particle term in Einstein's formula. 

As in the classical case, however, we are not done yet. Just as this does not seem to have occurred to Lorentz in the classical case, Jordan and his co-authors do not seem to have realized that one more step is needed to recover Einstein's formula. Eq.\ (\ref{q28}), like Eq.\ (\ref{22}), is for individual states, whereas what we need is a formula for a thermal ensemble of states. In quantum mechanics, this transition from individual states (pure states) to an ensemble of states (a mixed state) is a little trickier than in classical theory. Before we show how this is done, we 
want to make some comments about the interpretation of Eq.\ (\ref{q28}) that will make it very clear that this formula does not give the {\it thermal} mean square energy fluctuation. In modern terms,
the formula is for the mean square {\it quantum uncertainty} or {\it quantum dispersion} in the energy in a narrow frequency range in a small segment of the string when the whole string is in an energy eigenstate $\{ n_\nu \}$ . The operators $H$ and $H_{(a, \nu)}$ do not commute. The system is in an eigenstate of the full Hamiltonian $H$ but in a superposition of eigenstates of the Hamiltonian $H_{(a, \nu)}$ of the subsystem. Eq.\ (\ref{q28}) is a measure for the spread in the eigenvalues of the eigenstates of $H_{(a, \nu)}$ that make up this superposition rather than a measure of the spread in the value of the energy in the subsystem in an thermal ensemble of eigenstates of the system as a whole. It is that latter spread that gives the thermal mean square energy fluctuation. 

Proceeding as we did in the classical case (see Eqs.\ (\ref{23})--(\ref{30})), we make the transition from the formula for the mean square quantum uncertainty of the energy of the subsystem in an energy eigenstate of the whole system to the formula for the mean square fluctuation of this quantity in an ensemble of such states. As in the classical case, it turns out that these two formulae have the same form. The reason for this is once again that the entire
contribution to the mean square energy fluctuation in Eq.\ (\ref{q28}) comes from off-diagonal (i.e., $j \neq k$) terms involving the
product of two distinct, and therefore thermally uncorrelated, modes, as can clearly be seen, for instance, in Eq.\ (\ref{q30}). The thermal average of the product $H_j H_k$ is the product of the thermal averages of $H_j$ and $H_k$. This is a special feature of the system of uncoupled harmonic oscillators that we are considering and will not hold in general. In this special case, it turns out, we get from the formula for one state to the formula for a thermal ensemble of states simply by replacing the excitation levels $n_i$ in Eq.\ (\ref{q24}) by the thermal excitation levels given by the Planck function (cf.\ 3M, p.\ 379\footnote{The criticism of \citep{Debye 1910} at this point is retracted in \citep[p.\ 182, note]{Jordan 1928}.})
\begin{equation}
\hat{n}_{j} \equiv \frac{1}{e^{kT/h \nu_j}-1},
\label{q37}
\end{equation}
and repeat the steps that took us from Eq.\ (\ref{q24}) to Eq.\ (\ref{q28}).

We now show this in detail,  taking Eq.\ (\ref{q30}) as our starting point. We imagine the string to be coupled to an infinite external heat bath at temperature
$T$, with Boltzmann factor $\beta \equiv 1/kT$. The value of some observable in thermal equilibrium is given by the canonical-ensemble expectation value of the diagonal matrix elements of the corresponding operator $O$ in eigenstates $\{ n_i \}$ of the Hamiltonian for the system as a whole:
\begin{equation}
  \langle O(\{ n_i \}, \{ n_i \}) \rangle = \frac{ \displaystyle{ \sum_{ \{ n_i \} } } O(\{ n_i \}, \{ n_i \}) 
  e^{ \displaystyle{ -\beta E_{ \{ n_i \}}}}}{ \displaystyle{ \sum_{\{ n_i \}}} e^{ \displaystyle{ -\beta E_{\{ n_i \}}}}}
  \label{canens}
  \end{equation}
 where $E_{ \{ n_i \}} = \sum_{n_i} \left( n_i + \frac{1}{2} \right) \hbar \omega_i$ (see Eq.\  (\ref{q6})).\footnote{It does not matter for the ensemble average whether or not we include the zero-point energy in $E_{ \{ n_i \}}$, since the contributions from the zero-point energy  to numerator and  denominator are the same and cancel.} We calculate $\langle \overline{\Delta H_{(a, \omega)}^2} (\{ n_i \}, \{ n_i \}) \rangle$, the thermal average of the diagonal matrix elements of the operator $\overline{\Delta H_{(a, \omega)}^2}$ in the state $\{ n_i \}$. The only non-trivial part of this calculation is to determine the thermal average of the matrix elements $H_j H_k(\{ n_i \}, \{ n_i \})$ (with $j \neq k$). These matrix elements are given by (cf.\ Eq.\ (\ref{q24})):
\begin{equation}
H_j H_k (\{ n_i \}, \{ n_i \}) = \left( n_j + \frac{1}{2} \right) \hbar \omega_j  \left( n_k + \frac{1}{2} \right) \hbar \omega_k.
\label{q32}
\end{equation}
For the thermal average of this expression, we find, using Eq.\ (\ref{canens}):
\begin{equation}
\langle H_j H_k (\{ n_i \}, \{ n_i \}) \rangle =  \frac{ \displaystyle{ \sum_{ \{ n_{i} \} } }
\left(n_{j}+ \displaystyle{ \frac{1}{2}} \right)\hbar\omega_{j}
\left(n_{k}+ \displaystyle{ \frac{1}{2}} \right)\hbar\omega_{k}
e^{ \displaystyle{ -\beta E_{ \{ n_{i} \} }} }}{\displaystyle { \sum_{ \{ n_{i} \} } } 
e^{ \displaystyle{  -\beta E_{\{ n_{i} \}} } } }.
\label{q33} 
\end{equation}
The sum over all possible states $\{ n_i \}$ in the denominator can be written as a product of sums over all possible values of the excitation level $n_i$ for all modes $i$:
\begin{equation}
\sum_{ \{ n_{i} \}  } e^{ \displaystyle{ -\beta E_{ \{ n_{i} \} } } } 
= \prod_{i=1}^\infty \left( \sum_{n_i = 1}^\infty 
e^{ \displaystyle{ -\beta  \left(n_{i}+\frac{1}{2}\right)\hbar\omega_{i} } } \right).
\label{q33a}
\end{equation}
For all but the $j^{\rm th}$ and the $k^{\rm th}$ mode the $i^{\rm th}$  factor in the denominator cancels against an identical factor in the numerator. Eq.\ (\ref{q33}) thus reduces to a product of two factors of the same form, one for the $j^{\rm th}$ mode and one for the $k^{\rm th}$ mode. Consider the former:
\begin{equation}
\frac{ \displaystyle{ \sum_{n_{j}} }
\left(n_{j}+ \displaystyle{ \frac{1}{2}} \right)\hbar\omega_{j}
e^{-\beta(n_{j}+\frac{1}{2})\hbar\omega_{j}}}{\displaystyle {\sum_{n_{j}}}e^{-\beta (n_{j}+\frac{1}{2}) \hbar\omega_{j}}} = \frac{1}{2} \hbar \omega_j + 
\frac{ \displaystyle{ \sum_{n_{j}} }
n_{j} \hbar\omega_{j}
e^{-\beta n_{j}\hbar\omega_{j}}}{\displaystyle{\sum_{n_{j}}}e^{-\beta n_{j} \hbar\omega_{j}}}.
\label{q34}
\end{equation}
The expression in the denominator in the second term on the right-hand side is a geometric series, which we shall call $Z$:
\begin{equation}
Z \equiv  \displaystyle{\sum_{n_{j}}}e^{-\beta n_{j} \hbar\omega_{j}} = \frac{1}{1 - e^{-\beta \hbar\omega_j }}.
\label{q35}
\end{equation}
The fraction of the two sums in Eq.\ (\ref{q34}) is just minus the derivative of $\ln Z$ with respect to $\beta$. Eq.\ (\ref{q35}) allows us to write this as:
\begin{equation}
- \frac{d}{d\beta} \ln Z = - \frac{1}{Z} \frac{dZ}{d\beta} = 
- \left( 1 - e^{-\beta \hbar\omega_j} \right) \frac{- \hbar \omega_j e^{-\beta \hbar\omega_j} }{
\left( 1 - e^{-\beta \hbar\omega_j} \right)^2 } = \frac{ \hbar \omega_j}{ e^{\beta \hbar\omega_j} -1},
\label{q36}
\end{equation}
which is equal to $\hat{n}_j \hbar \omega_j$, where we used Eq.\ (\ref{q37}) for the thermal excitation levels. The right-hand side of Eq.\ (\ref{q34}) thus becomes $\left( \hat{n}_j + \frac{1}{2} \right) \hbar \omega_j$. Using this result for the $j^{\rm th}$ mode and a similar result for $k^{\rm th}$ mode, we can write Eq.\ (\ref{q33}) as
\begin{eqnarray}
\langle H_j H_k (\{ n_i \}, \{ n_i \}) \rangle & = & \langle H_j (\{ n_i \}, \{ n_i \}) \rangle
\langle H_k (\{ n_i \}, \{ n_i \}) \rangle
\nonumber \\
 & & \label{q38} \\
  & = & \left( \hat{n}_j + \frac{1}{2} \right) \hbar \omega_j
\left( \hat{n}_k + \frac{1}{2} \right) \hbar \omega_k \nonumber
\end{eqnarray}
Using this result, we calculate the thermal average of the diagonal matrix elements in the state $\{ n_i \}$ of the operator in Eq.\ (\ref{q30}) for the mean square energy fluctuation in a narrow frequency interval in a small segment of the string:
\begin{eqnarray}
\langle \overline{\Delta E_{(a, \omega)}^2} \rangle  & \equiv & \langle \overline{\Delta H_{(a, \omega)}^2} (\{ n_i \}, \{ n_i \}) \rangle \nonumber \\
& & \label{q39} \\ 
 & = & \frac{1}{l^2}  \sum_{j \neq k} \left(  \hat{n}_{j}\hat{n}_{k}+\frac{1}{2} \left( \hat{n}_{j}+\hat{n}_{k} \right)
 \right) \hbar^{2}\omega_{j}\omega_{k}  \frac{\sin^2{((\omega_j - \omega_k)a)}}{(\omega_j - \omega_k)^2}.  \nonumber
\end{eqnarray}
The right-hand side has exactly the same form as Eq.\ (\ref{q24}), except that the $n$'s are replaced by $\hat{n}$'s. Eq.\ (\ref{q37}) tells us that $\hat{n}_j$ and $\hat{n}_k$ vary smoothly with $j$ and $k$, so we can make the transition from sums to integrals in this case without any further assumptions. Proceeding in the exact same way as we did to get from Eq.\ (\ref{q24}) to Eq.\ (\ref{q21b}), we arrive at:
 \begin{equation}
\langle \overline{\Delta E_{(a, \nu)}^2} \rangle = 2 a \Delta \nu  \left( (\hat{n}_\nu h \nu)^2 + ( \hat{n}_\nu h \nu) \nu \right),
\label{q41}
\end{equation}
The thermal average of the mean excitation energy in a narrow frequency interval in a small segment of the string is given by (cf.\ 3M, Ch.\ 4, Eq.\ (39))
\begin{equation}
\langle \overline{E_{(a, \nu)} } \rangle = \frac{a}{l} \left( \hat{n}_\nu h \nu \right) N_\nu = 2 a \Delta \nu \left( \hat{n}_\nu h \nu \right),
\label{q42}
\end{equation}
This is just Eq.\ (\ref{q25}) with $\hat{n}$ instead of $n$.  With the help of this expression we can rewrite Eq.\ (\ref{q41}) as:
\begin{equation}
\langle \overline{\Delta E_{(a, \nu)}^2} \rangle =
\frac{ \langle \overline{ E_{(a, \nu)}} \rangle^2}{2a\Delta \nu} +\langle  \overline{E_{(a, \nu)}} \rangle h \nu.
\label{q43}
\end{equation}
This formula for the canonical-ensemble average of
the mean square fluctuation of the energy in a narrow frequency range in a small segment of the string has exactly the same form as Eq.\ (\ref{q28}) for the mean square quantum uncertainty in the energy of this subsystem in an energy eigenstate of the system as a whole. 

Our final result, Eq.\ (\ref{q43}), is the analogue for the simple model of a string of Einstein's famous 1909 formula for the mean square fluctuation of the energy in a narrow frequency range in a subvolume of a box with black-body radiation. That Eq.\ (\ref{q43}) emerges from the quantum-mechanical treatment of the modes of a string shows that the fluctuation formula, contrary to what Einstein thought, does not call for two separate mechanisms, one involving particles and one involving waves. In matrix mechanics, both terms arise from a single consistent dynamical framework. In the {\it Dreim\"annerarbeit} this unified mechanism is described in terms of quantized waves. If we focus on the occupation levels $n_i$ rather than on the field $u(x,t)$, however, we see that the same mechanism can also be described in terms of particles, quanta of the field, satisfying Bose's statistics. 

\section{Assessment of the validity and the importance of Jordan's argument}

The main conclusion we want to draw from our reconstruction of the fluctuation considerations in the {\it Dreim\"annerarbeit} is that they support the authors' claim---or rather Jordan's claim---that a straightforward application of the new matrix mechanics to a simple model of black-body radiation, viz.\ oscillations in a string fixed at both ends, leads to an expression for the mean square energy fluctuation in a narrow frequency range in a small segment of that string that has exactly the same form as the formula Einstein derived from thermodynamics and Planck's black-body radiation law for the mean square energy fluctuation in a narrow frequency range in a subvolume of a box filled with black-body radiation. We also noted, however, that the authors use a sloppy notation and that the argument they present is incomplete. 

At various points, the notation fails to reflect the crucial restriction to a narrow frequency range. We drew attention to a couple of passages in the text that clearly indicate that such a restriction is nonetheless in effect throughout the calculation. Since the entire derivation is for a finite frequency range, there are no problems with infinities ({\it pace} Ehlers,  2007, pp.\ 28--29). Another problem is that the authors do not distinguish in their notation between (in modern terms) operators and expectation values of operators in energy eigenstates. Here we have to keep in mind that this distinction had not fully crystalized when the paper was written. The authors had no clear notion yet of operators acting on states. They did not even have the general concept of a state \citep[sec.\ 3]{Duncan and Janssen}.

In the absence of the general state concept, they did not distinguish between pure states and mixed states either. This did trip them up. The formula they derived is for the mean square {\it quantum uncertainty} in the energy of a subsystem in an energy eigenstate of the system as a whole, which is a {\it pure state}. 
What they should have derived to recover Einstein's fluctuation formula is a formula for the {\it thermal} mean square fluctuation in the energy of the subsystem, i.e., a canonical-ensemble average over energy eigenstates of the whole system, which is a {\it mixed state}. We showed in detail how to make this transition from quantum uncertainty to thermal fluctuations. Given the preliminary character of the theory he was working with, Jordan can be forgiven for the omission of this step in the {\it Dreim\"annerarbeit}, though he probably should have known better when he presented his result again in later publications. In Jordan's defense,  we noted that Lorentz also omitted the corresponding step in the classical calculation.

With our admittedly not unimportant emendation, Jordan's result resolves Einstein's conundrum of the wave-particle duality of light. We cannot articulate the intrinsic value of the result any better than Jordan himself did in his contribution to the proceedings of a conference in Charkow in May 1929:
\begin{quotation}
Einstein drew the conclusion that the wave theory would necessarily have to be replaced or at least supplemented by the corpuscular picture. With our findings, however, the problem has taken a completely different turn. We see that it is not necessary after all to abandon or restrict the wave theory in favor of other models; instead it just comes down to reformulating the wave theory in quantum mechanics. The fluctuation effects, which prove the presence of corpuscular light quanta in the radiation field, then arise automatically as consequences of the wave theory.  The old and famous problem how one can understand waves and particles in radiation in a unified manner can thus in principle be considered as solved \citep[p.\ 702]{Jordan 1929}. 
\end{quotation}
Few of his colleagues shared Jordan's own assessment of the importance of his result. The main reason for its lukewarm reception in the physics community of his day seems to have been that the result looked suspicious because of the infinities one already encounters in this simple example of a quantum field theory. This suspicion has lingered, even though, as we saw, Jordan managed to steer clear of infinities by focusing on a narrow frequency range. 

The less than enthusiastic reaction of the physicists no doubt partly explains why Jordan's result has not become a staple of the historical literature on the wave-particle duality of light.  Another factor responsible for its neglect in this context, as we suggested in sec.\ 2, may have been that Jordan's result was too many things at once. It was the resolution of the conundrum of the wave-particle duality of light but it was also a striking piece of evidence for matrix mechanics and a telltale sign that a quantum theory of fields was needed.  Given how strongly Jordan felt about this last use of his result, it is perhaps only fitting that his derivation of Einstein's fluctuation formula has found its place in the historical literature not  toward the end of histories of wave-particle duality but at the beginning of histories of quantum field theory.  Still, the result only played a relatively minor role in the early stages of quantum mechanics and quantum field theory.  By contrast, it is the grand finale of the early history of the wave-particle duality of light. Regrettably, it has either been ignored in that context  or doubts have been cast upon it. We hope that our paper will help remove those doubts so that Jordan's result can finally be given its rightful place in the heroic tale of Einstein, light quanta, and the wave-particle duality of light.

\section*{Acknowledgments}
  
A preliminary version of this paper was presented at HQ1, a conference on the history of quantum physics held at the {\it Max-Planck-Institut f\"ur Wissenschaftsgeschichte}, Berlin, July 2--6, 2007. We are  grateful to Clayton Gearhart, Don Howard, Alexei Kojevnikov, Jos Uffink, and, especially, J\"{u}rgen Ehlers for helpful comments, discussion, and references. The research of Anthony Duncan is supported in part by the National Science Foundation under grant PHY-0554660.



\begin{thebibliography}{}

\bibitem[Bach(1989)]{Bach 1989} Bach, A.\ (1969). Eine Fehlinterpretation mit Folgen: Albert Einstein und der Welle-Teilchen Dualismus. {\it Archive for History of Exact Sciences} 40: 173--206.

\bibitem[Balashov and Vizgin(2002)]{Balashov and Vizgin 2002} Balashov, Y., and Vizgin, V.\  eds., {\it Einstein studies in Russia}. Boston: Birkh\"{a}user.

\bibitem[Baym(1969)]{Baym 1969} Baym, G.\ (1969). {\it Lectures on quantum mechanics.} Reading, MA: Addison-Wesley.

\bibitem[Beyler(2007)]{Beyler 2007} Beyler, R.\ H.\  (1969). Exporting the quantum revolution: Pascual Jordan's biophysical initiatives. In \citep[pp.\ 69--81]{Hoffmann 2007}

\bibitem[Bohr, Kramers, and Slater(1924a)]{BKS} Bohr, N., Kramers, H.\ A., and Slater, J.\ C.\  (1924a). The quantum theory of radiation {\it Philosophical Magazine} 47: 785--802. Reprinted in \citep[pp.\ 159--176]{Van der Waerden}.

\bibitem[Born(1925)]{Born 1925} Born, M.\ (1925). {\it Vorlesungen \"{u}ber Atommechanik.} Berlin: Springer.

\bibitem[Born(1978)]{Born 1978} Born, M.\ (1978). {\it My life. Recollections of Nobel laureate.} New York: Charles Scribner.

\bibitem[Born and Fuchs(1939a)]{Born and Fuchs 1939a} Born, M., and Fuchs, K.\ (1939a). On fluctuations in electromagnetic radiation. {\it Proceedings of the Royal Society of London, Series A, Mathematical and Physical
Sciences.} 170: 252--265.

\bibitem[Born and Fuchs(1939b)]{Born and Fuchs 1939b} Born, M., and Fuchs, K.\  (1939b). On fluctuations in electromagnetic radiation. (A correction.) {\it Proceedings of the Royal Society of London, Series A, Mathematical and Physical
Sciences.} 172: 465--466.

\bibitem[Born, Heisenberg, and Jordan(1926)]{dreimaenner} Born, M., Heisenberg, W., and Jordan, P.\ (1926). Zur Quantenmechanik II. {\it Zeitschrift f\"{u}r Physik}  35: 557--615. Page references to the English translation in \citep[pp.\ 321--385]{Van der Waerden}.

\bibitem[Born and Jordan(1925)]{Born and Jordan 1925} Born, M., and Jordan, P.\ (1925). Zur Quantenmechanik. {\it Zeitschrift f\"{u}r Physik} 34: 858--888. Page references are to the English translation of chs.\ 1--3 in \citep[pp.\ 277--306]{Van der Waerden}. Ch.\ 4 is omitted in this translation.

\bibitem[Born and Jordan(1930)]{Born and Jordan 1930} Born, M., and Jordan, P.\  (1930). {\it Elementare Quantenmechanik.} Berlin: Springer.

\bibitem[Bose(1924)]{Bose 1924} Bose, S.\ N.\ (1924). Plancks Gesetz und Lichtquantenhypothese. {\it Zeitschrift f\"{u}r Physik} 26: 178--181.

\bibitem[Bothe(1923)]{Bothe 1923} Bothe, W.\ (1923). Die r\"aumliche Energieverteilung in der Holhraumstrahlung. {\it Zeitschrift f\"ur Physik} 20: 145--152.

\bibitem[Bothe(1924)]{Bothe 1924} Bothe, W.\ (1924). \"Uber die Wechselwirkung zwischen Strahlung und freien Elektronen. {\it Zeitschrift f\"ur Physik} 23: 214--224.

\bibitem[Bothe(1927)]{Bothe 1927} Bothe, W.\ (1927). Zur Statistik der Hohlraumstrahlung. {\it Zeitschrift f\"ur Physik} 41: 345--351


\bibitem[Brush(2007)]{Brush 2007} Brush, S.\ G.\ (2007). How ideas became knowledge: The light-quantum hypothesis 1905--1935. {\it Historical Studies in the Physical and Biological Sciences} 37: 205--246.

\bibitem[Compton(1923)]{Compton 1923} Compton, A.\ H.\ (1923). A quantum theory of the scattering of X-rays by light elements. {\it Physical Review} 21: 483--502.

\bibitem[Darrigol(1986)]{Darrigol 1986} Darrigol, O.\ (1986). The origin of quantized matter waves. {\it Historical Studies in the Physical and Biological Sciences} 16: 197--253.

\bibitem[Debye(1910)]{Debye 1910} Debye, P.\ (1910). Der Wahrscheinlichkeitsbegriff in der Theorie der Strahlung. {\it Annalen der Physik} 33: 1427--1434.

\bibitem[Dirac(1927)]{Dirac 1927} Dirac, P.A.M.\ (1927).  The quantum theory of the emission and absorption of radiation. {\it Proceedings of the Royal Society of London, Series A, Mathematical and Physical Sciences.} 114: 243--265. Reprinted in \citep[pp.\ 1--23]{Schwinger 1958}

\bibitem[Duncan and Janssen(2007)]{Duncan and Janssen} Duncan, A., and Janssen, M.\  (2007). On the verge of {\it Umdeutung} in Minnesota: Van Vleck and the correspondence principle. Two parts. {\it Archive for History of Exact Sciences.} forthcoming.

\bibitem[Ehlers(2007)]{Ehlers 2007} Ehlers, J.\ (2007). Pascual Jordan's role in the creation of quantum field theory. In \citep[pp.\ 23--35]{Hoffmann 2007}. 


\bibitem[Ehrenfest(1925)]{Ehrenfest 1925a} Ehrenfest, P.\ (1925). Energieschwankungen im Strahlungsfeld oder Kristallgitter bei Superposition quantisierter Eigenschwingungen. 
{\it Zeitschrift f\"{u}r Physik} 34: 362--373.


\bibitem[Einstein(1905)]{Einstein 1905} Einstein, A.\ (1905). \"Uber einen die Erzeugung und Verwandlung des Lichtes betreffenden heuristischen Gesichtspunkt. {\it Annalen der Physik} 17: 132--148. Reprinted in facsimile as Doc.\ 14 in \citep[Vol.\ 2]{Einstein 1987--2006}. 

\bibitem[Einstein(1909a)]{Einstein 1909a} Einstein, A.\ (1909a). Zum gegenw\"artigen Stand des Strahlungsproblems. {\it Physikalische Zeitschrift} 10: 185--193. Reprinted in facsimile as Doc.\ 56 in \citep[Vol.\ 2]{Einstein 1987--2006}.

\bibitem[Einstein(1909b)]{Einstein 1909b} Einstein, A.\ (1909b). \"Uber die Entwickelung unserer Anschauungen \"uber das Wesen und die Konstitution der Strahlung. {\it Deutsche Physikalische Gesellschaft. Verhandlungen} 11: 482--500. Reprinted in facsimile as Doc.\ 60 in \citep[Vol.\ 2]{Einstein 1987--2006}.

\bibitem[Einstein(1914)]{Einstein 1914} Einstein, A.\ (1914). Zum gegenw\"artigen Stande des Problems der spezifischen W\"arme. Pp.\ 330--352 in: Arnold Eucken, ed., {\it Die Theorie der Strahlung und der Quanten. Verhandlungen auf einer von E.\ Solvay einberufenen Zusammenkunft (30.\ Oktober bis 3.\ November 1911), mit einem Anhange \"uber die Entwicklung der Quantentheorie vom Herbst 1911 bis Sommer 1913.} Halle a.S.: Knapp (Abhandlungen der Deutschen Bunsen Gesellschaft f\"ur angewandte physikalische Chemie, Vol.\ 3, no.\ 7). Reprinted in facsimile as Doc.\ 26 in \citep[Vol.\ 3]{Einstein 1987--2006}.



\bibitem[Einstein(1917)]{Einstein 1917} Einstein, A.\ (1917). Zur Quantentheorie der Strahlung. {\it Physikalische Zeitschrift} 18: 121--128. English translation in \citep[pp.\ 63--77]{Van der Waerden}.

\bibitem[Einstein(1925)]{Einstein 1925} Einstein, A.\ (1925). Bemerkung zu P.\ Jordans Abhandlung ``Zur Theorie der Quantenstrahlung." {\it Zeitschrift f\"{u}r Physik} 31: 784--785.



\bibitem[Einstein(1987--2006)]{Einstein 1987--2006} Einstein, A.\ (1987--2006). {\it The collected papers of Albert Einstein.} 10 Vols.\ Edited by J. Stachel {\it et al.\ } Princeton: Princeton University Press.

\bibitem[F\"olsing(1997)]{Foelsing 1997} F\"olsing, A.\ (1997). {\it Albert Einstein. A biography.} New York: Viking.

\bibitem[F\"urth(1928)]{Fuerth 1928} F\"urth, R.\ (1928). \"Uber Strahlungsschwankungen nach der Lichtquantenstatistik.  {\it Zeitschrift f\"{u}r Physik} 50: 310--318.


\bibitem[Gonzalez and Wergeland(1973)]{Gonzalez and Wergeland 1973} Gonzalez, J.\ J.\  and Wergeland, H.\  (1973). Einstein-Lorentz formula for the fluctuations of electromagnetic energy. {\it Det Kongelige Norske videnskabers selskab. Skrifter} 4: 1--4.


\bibitem[Heisenberg(1925)]{Heisenberg 1925} Heisenberg, W.\ (1925). \"{U}ber die quantentheoretische Umdeutung kinematischer und mechanischer Beziehungen. {\it Zeitschrift f\"{u}r Physik} 33: 879--893. Page references to English translation in \citep[pp.\ 261--276]{Van der Waerden}.


\bibitem[Heisenberg(1926)]{Heisenberg 1926b} Heisenberg, W.\ (1926). Schwankungserscheinungen und Quantenmechanik. {\it Zeitschrift f\"{u}r Physik} 40: 501--506. 

\bibitem[Heisenberg(1930)]{Heisenberg 1930} Heisenberg, W.\ (1930). {\it The physical principles of the quantum theory.} Chicago: University of Chicago Press.


\bibitem[Heisenberg(1931)]{Heisenberg 1931} Heisenberg, W.\ (1931). \"Uber Energieschwankungen in einem Strahlungsfeld. {\it Berichte \"uber die Verhandlungen der 
S\"achsischen Akademie der Wissenschaften zu Leipzig, mathematisch-physikalische Klasse} 83: 3--9. 


\bibitem[Heisenberg and Pauli(1929)]{Heisenberg and Pauli 1929} Heisenberg, W., and Pauli, W.\ (1929). Zur Quantendynamik der Wellenfelder. {\it Zeitschrift f\"{u}r Physik} 56: 1--61.

\bibitem[Heisenberg and Pauli(1930)]{Heisenberg and Pauli 1930} Heisenberg, W., and Pauli, W.\  (1930). Zur Quantendynamik der Wellenfelder, II. {\it Zeitschrift f\"{u}r Physik} 59: 168--190.

\bibitem[Hoffmann {\it et al.}(2007)]{Hoffmann 2007}  Hoffmann, D.,  Ehlers, J.,and Renn, J.\ eds.\ (2007). {\it Pascual Jordan (1902--1980). Mainzer Symposium zum 100.\ Geburtstag}. Max Planck Institute for History of Science, Preprint 329. 

\bibitem[Hoffmann and Walker(2007)]{Hoffmann and Walker 2007} Hoffmann, D., and Walker, M.\ (2007). Der gute Nazi: Pascual Jordan und das dritte Reich. In \citep[pp.\ 83--112]{Hoffmann 2007}.


\bibitem[Janssen(2006)]{Janssen 2006} Janssen, M., ed.\ (2006). {\it Studies in History and Philosophy of Modern Physics.} 37: 1--242 (Special issue---2005: The centenary of Einstein's {\it annus mirabilis}).

\bibitem[Jordan(1924)]{Jordan 1924} Jordan, P.\ (1924). Zur Theorie der Quantenstrahlung. {\it Zeitschrift f\"{u}r Physik} 30: 297--319.

\bibitem[Jordan(1925)]{Jordan 1925} Jordan, P.\  (1925). \"Uber das thermische Gleichgewicht zwischen Quantenatomen und Hohlraumstrahlung. {\it Zeitschrift f\"{u}r Physik} 33: 649--655.

\bibitem[Jordan(1927a)]{Jordan 1927a} Jordan, P.\  (1927a). Die Entwicklung der neuen Quantenmechanik. {\it Die Naturwissenschaften} 15: 614--623.

\bibitem[Jordan(1927b)]{Jordan 1927b} Jordan, P.\  (1927b). Die Entwicklung der neuen Quantenmechanik. Schlu\ss.  {\it Die Naturwissenschaften} 15: 636--649.

\bibitem[Jordan(1927c)]{Jordan 1927c} Jordan, P.\  (1927c). \"Uber Wellen und Korpuskeln in der Quantenmechanik. {\it Zeitschrift f\"{u}r Physik} 45: 766--775.

\bibitem[Jordan(1928)]{Jordan 1928} Jordan, P.\  (1928). Die Lichtquantenhypothese. Entwicklung und gegenw\"artiger Stand. {\it Ergebnisse der exakten Naturwissenschaften} 7: 158--208.

\bibitem[Jordan(1929)]{Jordan 1929} Jordan, P.\  (1929). Der gegenw\"artige Stand der Quantenelektrodynamik. {\it Physikalische Zeitschrift} 30: 700--713.

\bibitem[Jordan(1936)]{Jordan 1936} Jordan, P.\  (1936). {\it Anschauliche Quantentheorie. Eine Einf\"uhrung in die moderne Auffassung der Quantenerscheinungen.} Berlin: Julius Springer.

\bibitem[Jordan(1973)]{Jordan 1973} Jordan, P.\  (1973). Early years of quantum mechanics: some reminiscences. In \citep[pp.\ 294--299]{Mehra 1973}. 

\bibitem[Jordan and Klein(1927)]{Jordan and Klein 1927} Jordan, P., and Klein, O.\ (1927). Zum Mehrk\"orperproblem der Quantentheorie. {\it Zeitschrift f\"{u}r Physik} 45: 751--765.

\bibitem[Jordan and Pauli(1928)]{Jordan and Pauli 1928} Jordan, P., and Pauli, W.\ (1928). Zur Quantenelektrodynamik ladungsfreier Felder. {\it Zeitschrift f\"{u}r Physik} 47: 151--173.

\bibitem[Jordan and Wigner(1928)]{Jordan and Wigner 1928} Jordan, P., and Wigner, E.\ (1928). \"Uber das Paulische \"Aquivalenzverbot. {\it Zeitschrift f\"{u}r Physik} 47: 631--651.


\bibitem[Klein(1959)]{Klein 1959}  Klein, M.\ J.\ (1959). Ehrenfest's contributions to the development of quantum statistics. {\it Proceedings of the Koninklijke Nederlandse Akademie van Wetenschappen. Series B. Physical sciences} 62: 41--50, 51--62.

\bibitem[Klein(1964)]{Klein 1964}  Klein, M.\ J.\  (1964). Einstein and the wave-particle duality. {\it The Natural Philosopher} 3: 1--49.


\bibitem[Klein(1979)]{Klein 1979}  Klein, M.\ J.\  (1979). Einstein and the development of quantum physics. Pp.\ 133--151 in: A.\ P.\ French, ed., {\it Einstein. A centenary volume.} Cambridge, MA: Harvard University Press.  

\bibitem[Klein(1980)]{Klein 1980}  Klein, M.\ J.\  (1980). No firm foundation: Einstein and the early quantum theory. In \citep[pp.\ 161--185]{Woolf 1980}.

\bibitem[Klein(1982)]{Klein 1982}  Klein, M.\ J.\ (1982). Fluctuations and statistical physics in Einstein's early work. Pp.\ 39--58 in: G.\ Holton and Y.\ Elkana, eds, {\it Albert Einstein. Historical and cultural perspectives}. Princeton: Princeton University Press. (Reprinted: New York: Dover,1992.)

\bibitem[Kojevnikov(1990)]{Kojevnikov 1990}  Kojevnikov, A.\ (1990). Einshteinovskaya formula fluktuatsii i korpuskulyarno-volnovoy dualizm. Pp.\ 49--97 in: I.\ Yu.\ Kobzarev and G.\ E.\ Gorelik, eds., {\it Einshteinovskiy Sbornik, 1986--1990.} Moscow: Nauka. Translation, Einstein's fluctuation formula and the wave-particle duality, in \citep[pp.\ 181-228]{Balashov and Vizgin 2002}.

\bibitem[Kuhn {\it et al.}(1967)]{Kuhn et al. 1967} Kuhn, T.\ S., Heilbron, J.\ L., Forman, P., and Allen, L.\ (1967). {\it Sources for the history of quantum physics. An inventory and report.} Philadelphia: American Philosophical Society.

\bibitem[Kundt(2007)]{Kundt 2007}  Kundt, W.\ (2007). Jordan's ``excursion" into geophysics. In \citep[123--131]{Hoffmann 2007}



\bibitem[Laue(1915)]{Laue 1915c} Laue, M.\ von (1915). Die Einsteinschen Energieschwankungen. {\it Verhandlungen der Deutschen Physikalischen Gesellschaft} 17: 198--202.

\bibitem[Lorentz(1916)]{Lorentz 1916} Lorentz, H.\ A.\ (1916). {\it Les th\'eories statistiques en thermodynamique: conf\'erences faites au Coll\`ege de France en novembre 1912.} Leipzig, Berlin: Teubner.

\bibitem[Mehra(1973)]{Mehra 1973} Mehra, J., ed.\ (1973).  {\it The physicist's conception of nature.} Dordrecht: Reidel. 

\bibitem[Mehra and Rechenberg(1982--2001)]{Mehra Rechenberg} Mehra, J., and Rechenberg, H.\ (1982--2001). {\it The historical development of quantum theory.} 6 Vols.\ New York, Berlin: Springer.

\bibitem[Miller(1994)]{Miller 1994} Miller, A.\ I.\ (1994). {\it Early quantum electrodynamics. A source book.} Cambridge: Cambridge University Press.

\bibitem[Milonni(1981)]{Milonni 1981}  Milonni, P.\ W.\ (1981). Quantum mechanics of the Einstein-Hopf model. {\it American Journal of Physics} 49: 177--184.

\bibitem[Milonni(1984)]{Milonni 1984}  Milonni, P.\ W.\ (1984). Wave-particle duality of light: A current perspective. Pp.\ 27--67 in:  S. Diner, D.\ Fargue, G.\ Lochak, and F.\ Selleri, eds., {\it The wave-particle dualism: A tribute to Louis de Broglie on his 90th birthday} Dordrecht: Reidel.

\bibitem[Norton(2006)]{Norton 2006} Norton, J.\ D.\ (2006). Atoms, entropy, quanta: Einstein's miraculous argument of 1905. In \citep[71--100]{Janssen 2006}.


\bibitem[Pais(1979)]{Pais 1979} Pais, A.\ (1979). Einstein and the quantum theory. {\it Reviews of Modern Physics} 51: 863--914.

\bibitem[Pais(1980)]{Pais 1980} Pais, A.\ (1980). Einstein on particles, fields, and the quantum theory. In \citep[pp.\ 197--251]{Woolf 1980}.

\bibitem[Pais(1982)]{Pais 1982} Pais, A.\ (1982). {\it `Subtle is the Lord \ldots' The science and the life of Albert Einstein.} Oxford: Oxford University Press.

\bibitem[Pais(1986)]{Pais 1986} Pais, A.\ (1986). {\it Inward bound. Of matter and forces in the physical world.} Oxford: Clarendon.

\bibitem[Pauli(1926)]{Pauli 1926} Pauli, W.\ (1926). \"Uber das Wasserstoffspektrum vom Standpunkt der neuen Quantenmechanik.  {\it Zeitschrift f\"ur Physik} 36: 336--363. English translation in \citep[pp.\ 387--415]{Van der Waerden}.

\bibitem[Pauli(1930)]{Pauli 1930} Pauli, W.\ (1930). Review of \citep{Born and Jordan 1930}. {\it Die Naturwissenschaften} 18: 602.

\bibitem[Pauli(1949)]{Pauli 1949} Pauli, W.\ (1949). Einstein's contributions to quantum theory. Pp. 147--160 in: P.\ A.\ Schilpp, ed., {\it Albert Einstein: philosopher--scientist.} Evanston, IL: The Library of Living Philosophers.

\bibitem[Pauli(1979)]{Pauli 1979} Pauli, W.\ (1979). {\it Scientific correspondence with Bohr, Einstein, Heisenberg a.o. Volume I: 1919--1929.} Edited by A.\ Hermann, K.\ von Meyenn, and  V.\ F.\ Weisskopf. Berlin: Springer.

\bibitem[Peierls(1973)]{Peierls 1973} Peierls, R.\ E.\ (1973). The development of quantum field theory. In \citep[pp.\ 370--379]{Mehra 1973}.


\bibitem[Rynasiewicz and Renn(2006)]{Ryno-Renn 2006} Rynasiewicz, R., and Renn, J.\ (2006). The turning point for Einstein's {\it annus mirabilis}. In \citep[5--35]{Janssen 2006}.

\bibitem[Schroer(2007)]{Schroer 2007} Schroer, B.\ (2007). Pascual Jordan, biographical notes, his contributions to quantum mechanics and his role as a protagonist of quantum field theory. In \citep[pp.\ 47--68]{Hoffmann 2007}. 

\bibitem[Schweber(1994)]{Schweber 1994} Schweber, S.\ S.\ (1994). {\it QED and the men who made it: Dyson, Feynman, Schwinger, and Tomonaga.} Princeton: Princeton University Press.

\bibitem[Schwinger(1958)]{Schwinger 1958} Schwinger, J., ed.\ (1958). {\it Selected papers on quantum electrodynamics.} New York: Dover.

\bibitem[Schwinger(1973)]{Schwinger 1973} Schwinger, J.\ (1973). A report on quantum electrodynamics. In \citep[pp.\ 412--413]{Mehra 1973}.

\bibitem[Smekal(1926)]{Smekal 1926} Smekal, A.\ (1926). Zur Quantenstatistik der Hohlraumstrahlung und ihrer Wechselwirkung mit der Materie. {\it Zeitschrift f\"ur Physik} 37: 319--341. 

\bibitem[Stachel(1986)]{Stachel 1986} Stachel,  J.\ (1986). Einstein and the quantum. Fifty years of struggle. Pp.\ 349--385 in: R.\ G.\ Colodny, {\it From quarks to quasars. Philosophical problems of modern physics.} Pittsburgh: University of Pittsburgh Press. Page references to reprint in \citep[pp.\ 367--402]{Stachel 2002}.

\bibitem[Stachel(2002)]{Stachel 2002} Stachel,  J.\ (2002). {\it Einstein from `B' to `Z'}. Boston: Birkh\"{a}user.



\bibitem[Uffink(2006)]{Uffink 2006} Uffink,  J.\ (2006). Insuperable difficulties: Einstein's statistical road to molecular physics. In \citep[36--70]{Janssen 2006}.

\bibitem[Van der Waerden(1968)]{Van der Waerden} Van der Waerden, B.\ L., ed.\ (1968). {\it Sources of quantum mechanics.} New York: Dover.

\bibitem[Van Dongen(2007)]{Van Dongen 2007} Van Dongen, J.\ (2007). Emil Rupp, Albert Einstein, and the canal ray experiments on wave-particle duality: Scientific fraud and theoretical bias. {\it Historical Studies in the Physical and Biological Sciences} 37 (Supplement): 73--120.

\bibitem[Von Meyenn(2007)]{Von Meyenn 2007} Von Meyenn, K.\  (2007). Jordan, Pauli, und ihre fr\"uhe Zusammenarbeit auf dem Gebiet der Quantenstrahung. In \citep[pp.\ 37--46]{Hoffmann 2007}. 




\bibitem[Weinberg(1995)]{Weinberg 1995} Weinberg,  S.\ (1995). {\it The quantum theory of fields.} Vol.\ 1. {\it Foundations.} Cambridge: Cambridge University Press.

\bibitem[Wentzel(1960)]{Wentzel 1960} Wentzel, G.\ (1960). Quantum theory of fields (until 1947). Pp.\ 48--77 in: M.\ Fierz and V.\ F.\ Weiss\-kopf (eds.), {\it Theoretical physics in the twentieth century. A memorial volume to Wolfgang Pauli.} New York: Interscience Publishers. Reprinted in \citep[pp.\ 380--403]{Mehra 1973}.


\bibitem[Wolfke(1921)]{Wolfke 1921} Wolfke, M.\ (1921). Einsteinsche Lichtquanten und r\"aumliche Struktur der Strahlung. {\it Physikalische Zeitschrift} 22: 375--379.

\bibitem[Woolf(1980)]{Woolf 1980} Woolf, H., ed.\ (1980). {\it Some strangeness in the proportion. A centennial symposium to celebrate the achievements of Albert Einstein.} Reading, MA: Addison-Wesley.

\end{thebibliography}
\end{document}